\documentclass[fleqn,usenatbib]{mnras}

\usepackage{newtxtext,newtxmath}

\usepackage[T1]{fontenc}
\usepackage{xcolor}
\usepackage[normalem]{ulem} 

\DeclareRobustCommand{\VAN}[3]{#2}
\let\VANthebibliography\thebibliography
\def\thebibliography{\DeclareRobustCommand{\VAN}[3]{##3}\VANthebibliography}

\usepackage{graphicx}	
\usepackage{amsmath}	

\title[
Effects of flares on solar $p$-mode amplitudes]
{Effects of flares on solar high-degree helioseismic acoustic mode amplitudes}

\author[M. C. Rabello Soares et al.]{M. Cristina Rabello Soares,$^{1}$\thanks{E-mail: cristina@fisica.ufmg.br (MCRS)}
Fr\'ed\'eric Baudin,$^{2}$
and Vanessa G. Teixeira$^{1}$ 
\\
$^{1}$Physics Department, Universidade Federal de Minas Gerais, Belo Horizonte MG 31270-901, Brazil\\
$^{2}$Institut d'Astrophysique Spatiale, Universite Paris-Saclay Faculte des Sciences d'Orsay, F-91405 Orsay, France
}

\date{Accepted XXX. Received YYY; in original form ZZZ}

\pubyear{2021}

\begin{document}
\label{firstpage}
\pagerange{\pageref{firstpage}--\pageref{lastpage}}
\maketitle

\begin{abstract}
Several attempts have been made to observe whether solar flares excite acoustic modes
since 
Charles Wolff
suggested this possibility almost fifty years ago.
We look for the impact of flares on the amplitude of solar acoustic modes and other effects that are also affecting the mode amplitude. 
Solar acoustic mode amplitudes are known to be sensitive to magnetic fields.
As flares usually occur in the presence of strong magnetic fields and most likely are the by-product of magnetic reconnection, we show how the magnetic field in and around the flaring region affects the mode amplitude.
The mode amplitudes were obtained using ring-diagram analysis, which was first applied to a single event, the largest flare in the space age (the `Halloween Flare', SOL2003-10-28T11:00), using MDI data.
Then, using HMI data, the analysis was applied to the regions corresponding to the flares observed during the high activity phase of cycle 24 and that fall into two groups. These two groups consist of small (10-60 erg\,cm$^{-2}$~s$^{-1}$) and large ($>$1200 erg\,cm$^{-2}$~s$^{-1}$) peak-flux flares, based on the Heliophysics Event Knowledgebase (HEK). 
The first set is used as a comparison to the results of the strong flares in the second set.
After applying several corrections in order to take into account several
sources of bias, we did not find any amplification in the inferred mode
amplitude due to flaring activity
larger than
a 10\,\% uncertainty. 

\end{abstract}

\begin{keywords}
Sun: helioseismology -- Sun: flares -- Sun: activity
\end{keywords}



\section{Introduction}

Although we now know that the acoustic modes of the quiet Sun are excited by convection, the discussion by \citet{Wolff_1972_1} of the possible excitation of normal modes of solar oscillation ($p$-modes) by solar flares stimulated attempts to detect this excitation \citep{Haber_1988_1}. Possible correlations with CMEs were investigated \citep{Foglizzo_1998_2} as well as correlations between low-$\ell$ oscillations and soft X-rays \citep{Karoff_2008_1}, and even more recently by \citet{Kumar17}.
The rapid progress of asteroseismology and the study of stellar flares makes the study of these phenomena in the Sun important to inform our study of the influence of the more energetic stellar flares on asteroseismic acoustic modes.

Oscillations of small spatial scales were also investigated after the illustration of the effects of a flare on the photosphere \citep{Kosovichev_1998_1} and the  work of \citet{Donea_1999_3}.
Changes in the energy in
high-$\ell$ $p$-modes following flares were observed by \citet{maurya2009}. \citet{Maurya_2014} performed a statistical study on 53 events and found a different behaviour of $p$-mode parameters (amplitude, width) in flaring regions compared to active but non-flaring regions. 
However, identifying correlations between the temporal behaviour of $p$-modes and 
indicators of flares (or other violent events)
has led to contradictory results;for example, \citet{Richardson_2012_1} did not find a correlation between high-frequency $p$-modes and flares.\\
Indeed, mode amplitudes are very sensitive to
observational biases 
or problems, such as gaps in the data, which may
may alter the result. 
First, mode amplitudes
decrease approximately linearly with the mean magnetic field in the region where the waves are observed
\citep{Komm_2000,Rajaguru_2001,Howe_2004_2,Rabello-Soares_2008}. Then, when analysing a solar region before and after a flare, one needs to be aware of changes in the overall magnetic field of the sunspot in the region, which in general has a large area and strong magnetic fields. These changes make the amplitude behaviour of the modes more complex to interpret. Other biases also alter the observed amplitudes, such as the effect of the line of sight and gaps in the data. We list the main pitfalls that could be encountered in this analysis in Section\,\ref{sec:pitfalls}. 
In Section\,\ref{sec:results}, we analyse a single event, the "Halloween" flare.
Then, we analyse 
flares that occurred during the four years of the highest activity during cycle 24 (June 2012\,--\,August 2016) and fell in one of two groups: 
all flares with peak-flux larger than 1200 erg\,cm$^{-2}$~s$^{-1}$ (including GOES classes M and X)
and
another set with peak-flux between 10--60 erg\,cm$^{-2}$~s$^{-1}$
used as a comparison to the results of the strong flares set.
The conclusions are summarised in Section\,\ref{sec:conclusions}.\\
We focus here on the case of intermediate or high-degree modes; the case of low degree modes will be investigated in a future article.

\section{Pitfalls}
\label{sec:pitfalls}

We review here the main sources of bias when observing mode amplitude variation in time, which can interfere with intrinsic amplitude variation due to a possible excitation by flares.

\subsection{Influence of line of sight}
\label{sec:line-of-sight}

\begin{figure}
	\includegraphics[width=\columnwidth]{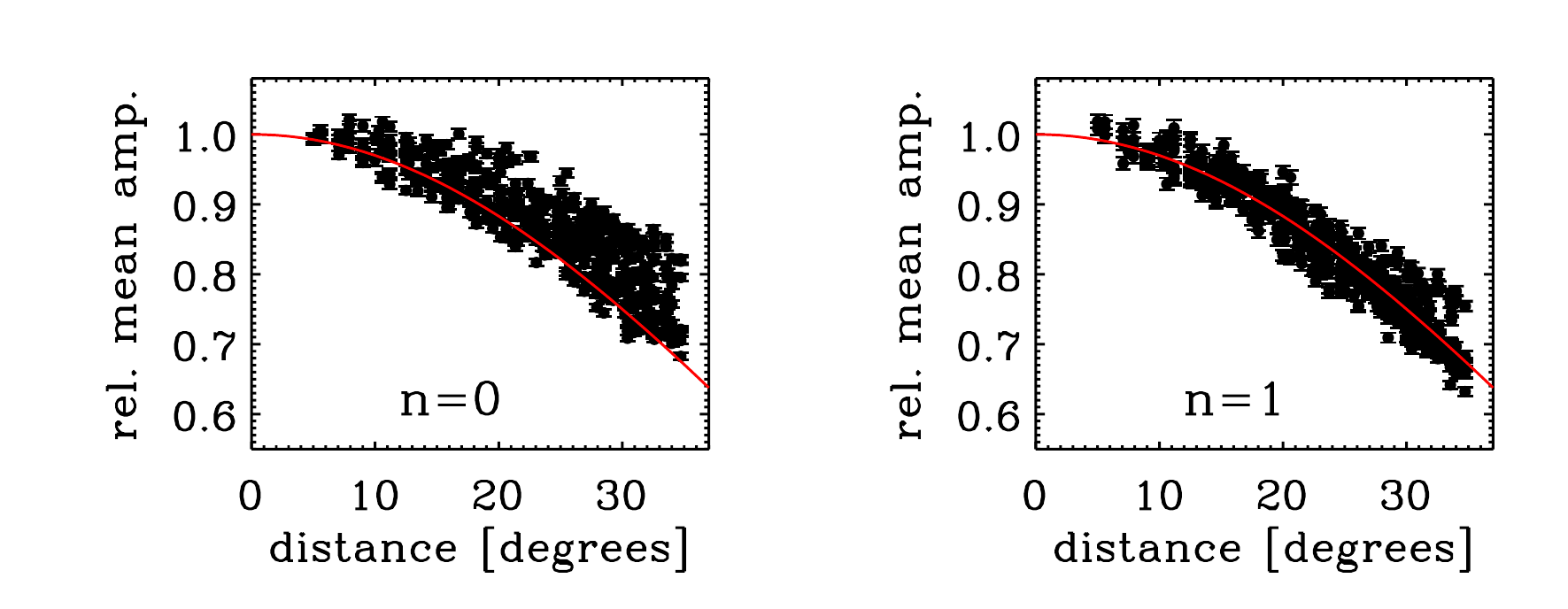}
    \caption{
    The mean mode amplitude divided by its maximum, as function of
    distance from disc centre for two acoustic modes with frequency $\nu = 3300\,\mu$Hz and radial order $n=0$ (left) and $n=1$ (right).
    The error bars are errors of the mean of 20 regions at each disc position.
    Mode amplitudes obtained from the HMI Ring Analysis pipeline for five-degree
    magnetically-quiet 
    tiles from August 2015 to August 2016 and used in Section\,\ref{sec:stat}.
   The red line is the square of the cosine of the distance from disc centre (i.e., the l.o.s.), considering that the y axis is the amplitude of the power spectra.
   }
    \label{fig:freq_bias_cris}
\end{figure}

\begin{figure}
	\includegraphics[width=\columnwidth]{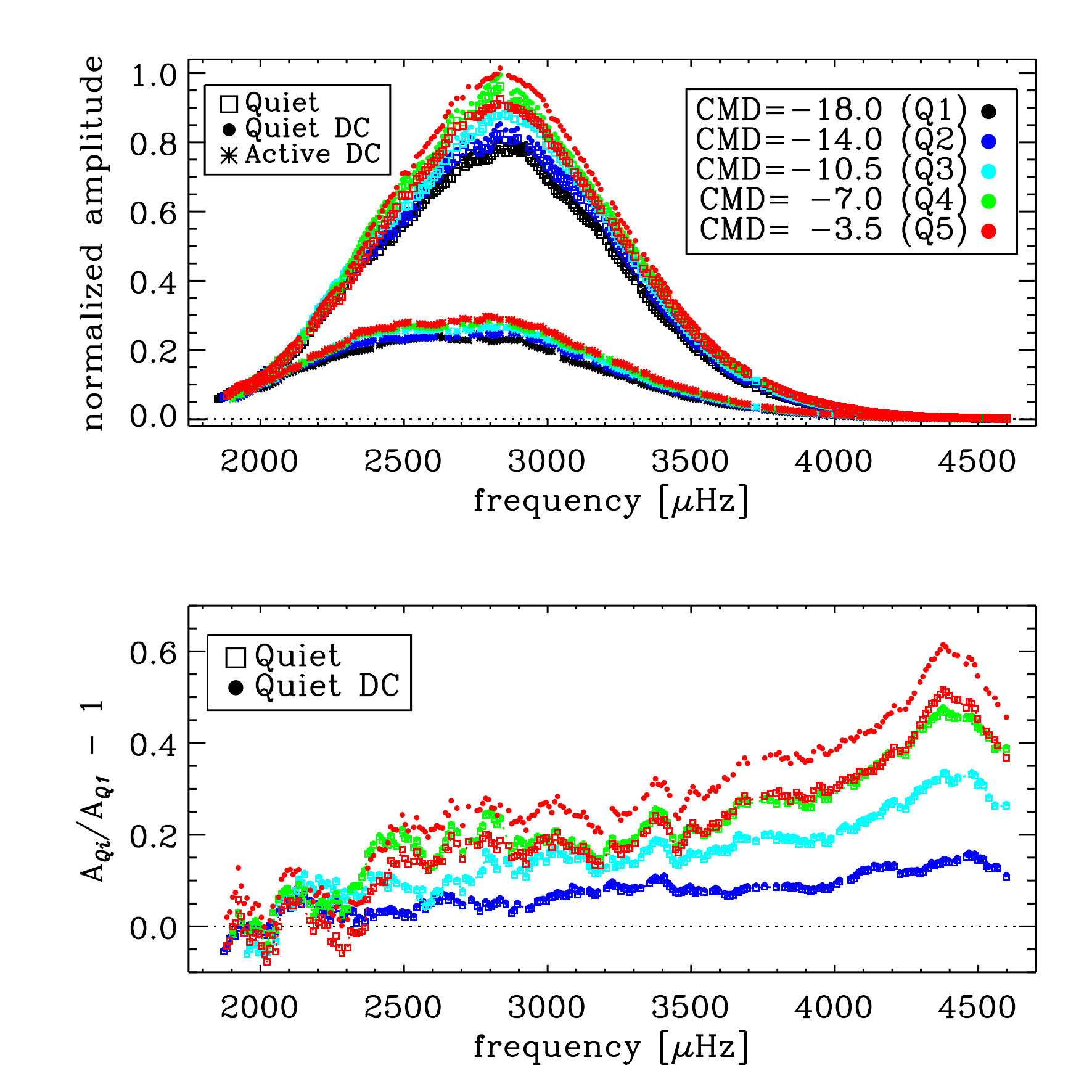}
    \caption{
    Top. Example of the normalized mode amplitude variation with frequency for modes with $n$=1 obtained at 16-degree MDI regions
    at latitude 22.5$\degr$ South and at five different CMD: 3.5 (red), 7 (green), 10.5 (light blue), 14 (dark blue) and 18$\degr$ (black), named Q1\,--\,5. As the Sun rotates, a region would move from 18$\degr$ to 3.5$\degr$ East. The squares are for quiet regions (MAI$\sim$20\,G) obtained during 25\,--\,26 October 2003. The full circles are for these quiet regions corrected for low data coverage, `DC' (Section\,\ref{sec:gaps}). The stars are active regions (MAI$\sim$270\,G) obtained during 27\,--\,28 October 2003, and they are corrected for reduced data coverage.
    The amplitude fitting uncertainties are smaller than 0.03 and approximately 5\,\% of the amplitude.
    Bottom. Relative mode amplitude in relation to those for one of the disc positions (Q1, CMD=18$\degr$ East) for the quiet regions with (full circles) and without (squares) the correction for the
    reduced
    data coverage. 
    }
    \label{fig:mdi_r15_los}
\end{figure}

The observed flaring region is carried by rotation across the Sun and is thus observed at
a varying angle (for a given latitude). 
The effect of the position on the disc can be clearly seen when considering the mode amplitudes at all frequencies across all of the positions on the disc when ring diagrams are calculated during a magnetically quiet day.
At a given centre-to-limb distance, the observed amplitude varies with frequency, $\nu$, and radial order $n$.
Figure~\ref{fig:freq_bias_cris} shows the variation of the mean mode amplitude with the distance of the observed region from the disc centre (averaged over 20 regions at each disc position) with a 100\,\% duty cycle for two different modes.
For these modes, their
amplitudes
decrease by
as much as $\sim$30\,\%
with distance from disc centre.
The modes observed in ring diagrams are not purely radial, so a simple correction taking into account the projection on the light-of-sight (l.o.s.) by dividing by the cosine of the angle between this l.o.s. and the radial direction is not sufficient.
The red line in Figure~\ref{fig:freq_bias_cris} corresponds to $\cos^2$(distance),
as the mode amplitude is the amplitude of the power spectrum.
The agreement is not as good for modes with $n=0$, 
where the ratio of the horizontal and radial components of the displacement is one.
%
%
%
The mode amplitude
also varies with latitude and central meridian distance (CMD), not only distance.
This is shown in the Figure as the amplitude variation ($\la$10\%) at a given distance.
This effect on helioseismic analysis has been known for sometime 
\citep[e.g.,][]{schou1998}
and research on this subject has been done by \citet{zhao2012, baldner_schou2012}.
Figure~\ref{fig:mdi_r15_los} shows the variation of mode amplitudes
with frequency for five different 16-degree regions with different CMD (squares: quiet regions)
at a given latitude, where we can see a significant change in the mode amplitude.

To take into account this effect, one can use a reference region, at the same position on the disc but observed when no activity at all is seen \citep[as in][]{Basu_2004}. Using the amplitudes of the modes derived from this quiet region at the same location, it is possible to estimate the amplitude variation in the observed active region by dividing them by those of the quiet region, both amplitudes being affected in the same way by l.o.s. effects. 
This method was adopted in this paper as described in Section\,\ref{sec:results}.

\subsection{Influence of MAI}
\label{sec:mai}

The reduction of the amplitude of acoustic modes in active regions,
approximately linearly with increase in the mean unsigned magnetic field,
is a well-known phenomena
\citep[\textit{e.g.}][]{Komm_2000,Rajaguru_2001,Howe_2004_2,Rabello-Soares_2008}.
If an increase in the mode amplitudes as a signature of a flare is investigated, the
variation due to magnetic damping 
could be an important
factor of variation, in particular in the case of a decrease of the magnetic field in the observed region,
due, for example, to sunspot evolution and the magnetic reconnection during the flare.

In Figure~\ref{fig:mdi_r15_los} (top panel),
we show active regions (stars) at different disc positions
that have their amplitude strongly reduced by the strong magnetic fields present,
in comparison to quiet regions
at the same disc position but a few days apart.
The magnetic activity index (MAI) is calculated by the HMI pipeline,
by averaging all pixels in the MDI 
magnetogram with a flux greater than 50\,G over the tracking interval \citep{Basu_2004,bogart2011a}.
In the Figure, the MAI varies, on average, from $\sim$12\,G to $\sim$270\,G from quiet to active regions.
The active regions in the figure corresponds to 
the archetypal Halloween flare 
which will be analysed in Section\,\ref{sec:Halloween} and 
named regions R1\,--\,5.
The MAI varied only
a few percent during the entire sequence R1\,--\,5.
As seen in Figure\,\ref{fig:mdi_r15_los}, amplitudes in the Halloween flare region are decreased by almost a factor of three, 
as initially observed by \citet{Braun_1987}. A small change in the magnetic field strength, inclination or even a change in the active region magnetic configuration will change the mode amplitude,
which could be confused as an effect of a flare.

\begin{figure}
	\includegraphics[width=\columnwidth]{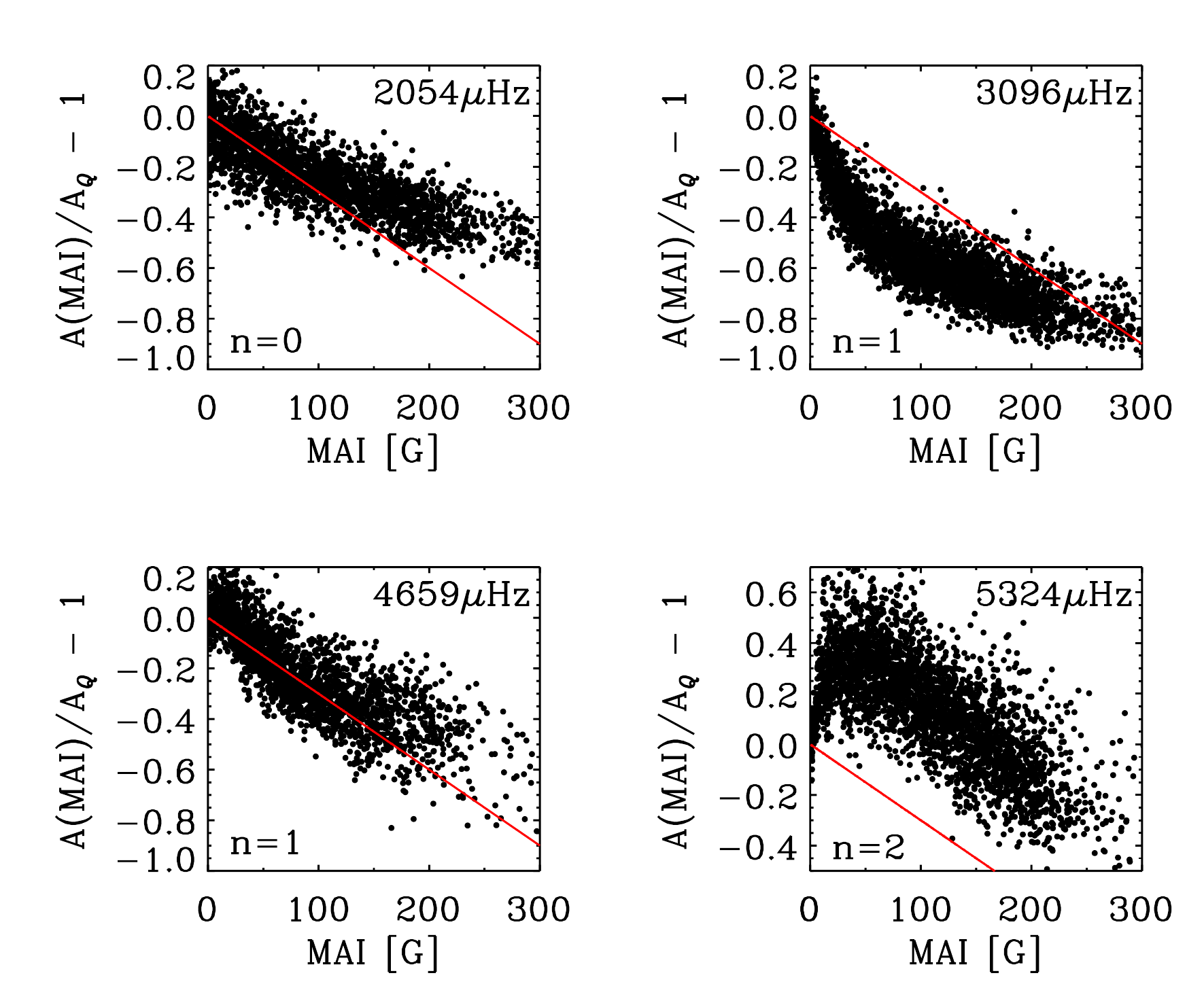}
   \caption{
   Variation of mode amplitude 
   with MAI for four different modes (clockwise): 
   ($n=0$ and $\nu$ = 2054 $\mu$Hz), 
   ($n=1$ and $\nu$ =  3096 $\mu$Hz),
   ($n=2$ and $\nu$ =  5324 $\mu$Hz), and
   ($n=1$ and $\nu$ =  4659 $\mu$Hz).
   Mode amplitudes were obtained from the HMI Ring Analysis pipeline for five-degree tiles for 20 quiet regions (MAI < 1\,G) at the same disc position, so that there was no effect of a changing line of sight. The increase of mode amplitude at high frequency (known as the acoustic halo effect) is seen in the bottom-right panel.
   The red line corresponds to coefficient of 
   0.003\,G$^{-1}$ given by \citet{Howe_2004_2} averaging all modes with frequency smaller than 4500 $\mu$Hz.
    \label{fig:amplirel_mai}
    }
\end{figure}

\citet{Rajaguru_2001} found a change of mode amplitude around -0.004\,G$^{-1}$, and \citet{Howe_2004_2} reported -0.003\,G$^{-1}$ for frequencies below 4500 $\mu$Hz.
A decrease in a region's MAI, as large as the Halloween Flare, of 10\,\% would imply an increase
of 11\,\% in the mode amplitude. 
Figure~\ref{fig:amplirel_mai} shows the variation with MAI of mode amplitude corrected for the line-of-sight effect for four different modes, 
where an average of 20 quiet regions (with MAI < 1\,G) at each disc position (used in Section\,\ref{sec:stat}) was applied as reference, $A_Q$.
For comparison, a line indicates a coefficient of -0.003\,G$^{-1}$ 
to emphasize the mode dependence.
The mode amplitude can vary by as much as 90\,\% in the presence of active regions. For high-frequency modes ($\nu$ > 5200 $\mu$Hz), there is an enhancement of the mode amplitude rather than an absorption in active regions: this is known as the acoustic halo effect
\citep{brown1992}, as seen in the bottom right panel.

\subsection{Gaps}
\label{sec:gaps}

\begin{figure}
   \centering
      \includegraphics[width=0.9\columnwidth]{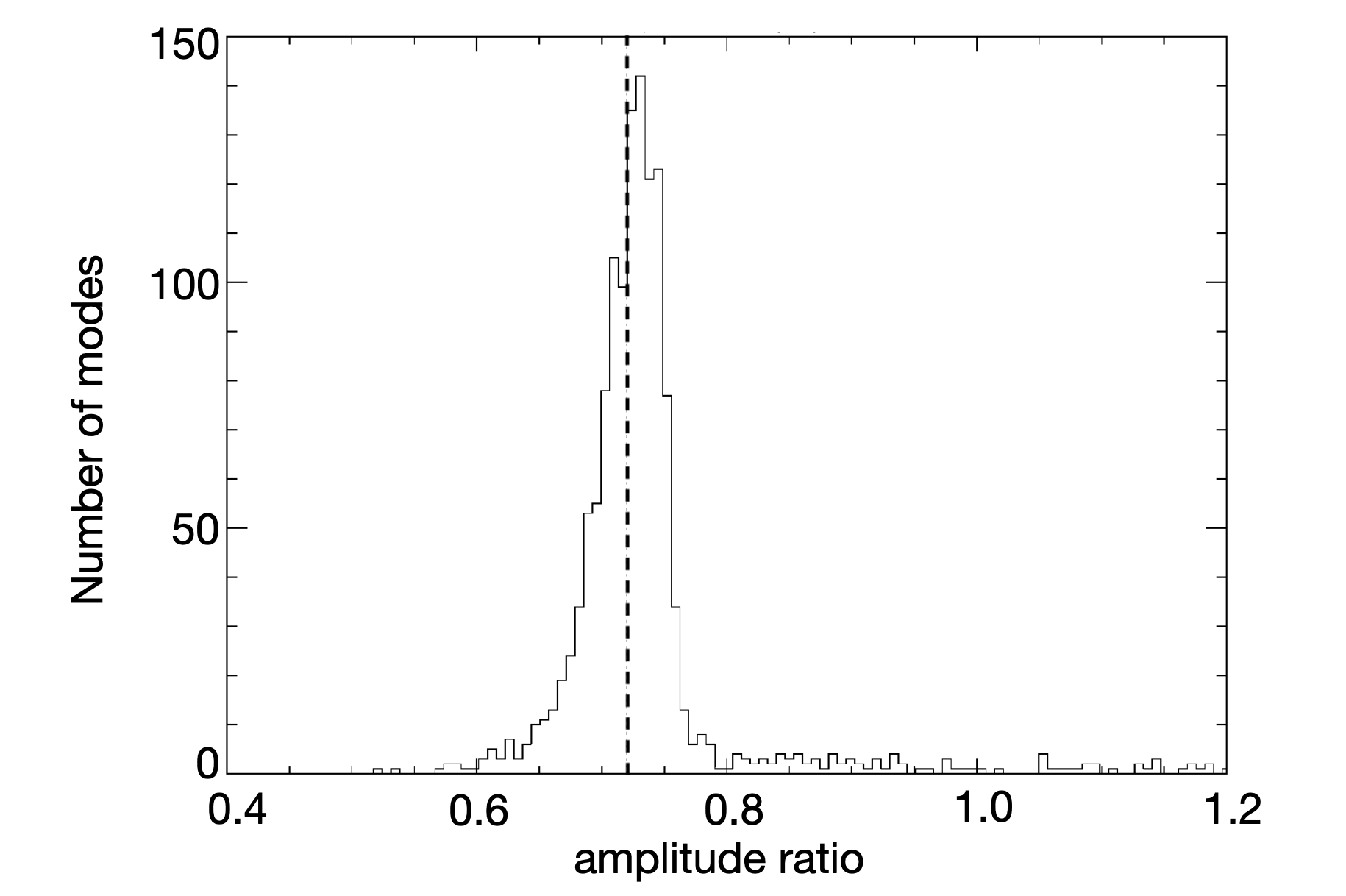}
      \includegraphics[width=0.9\columnwidth]{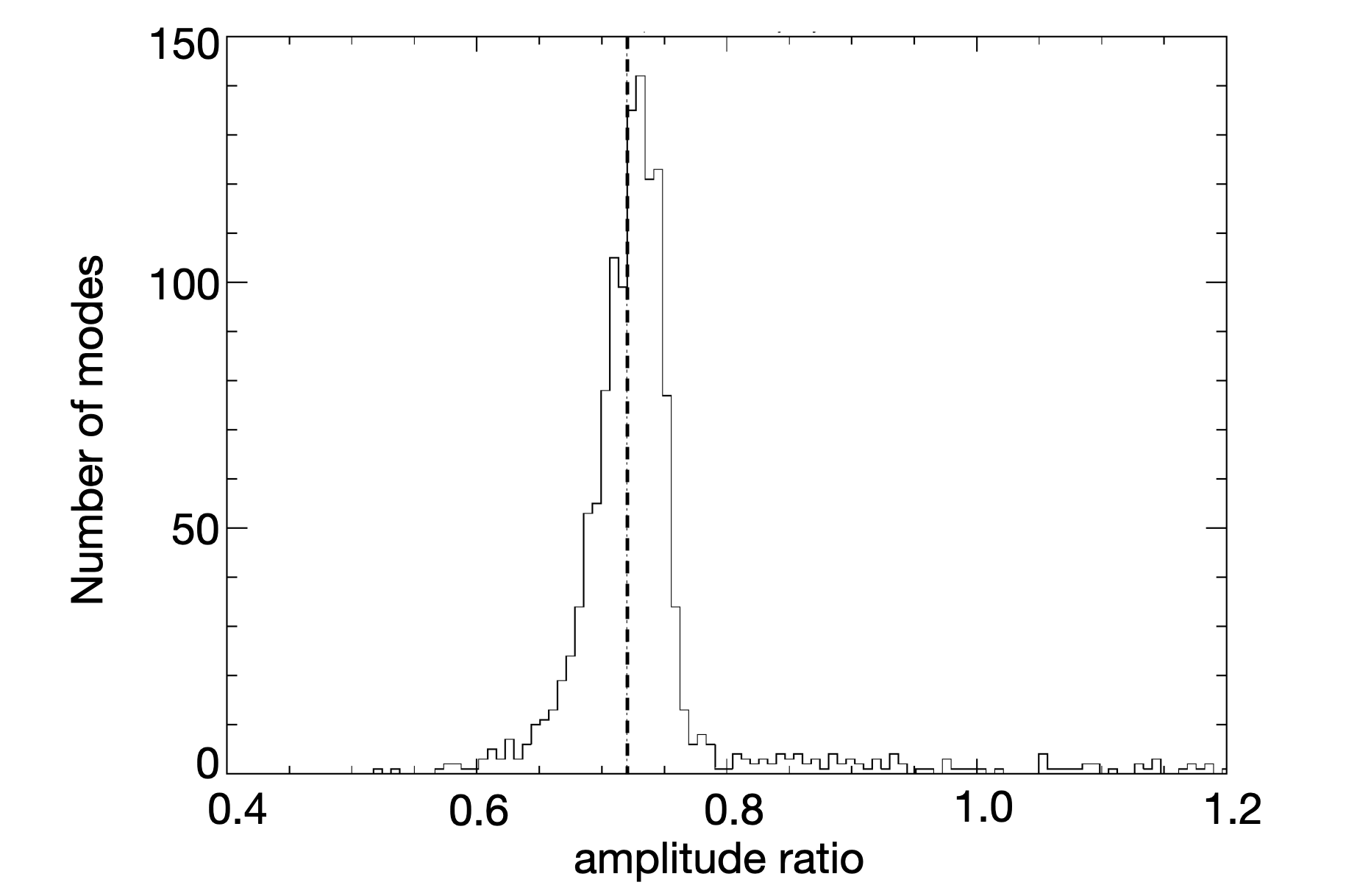}
      \caption{Histogram of the ratio of fitted amplitudes of 98\,\% (upper) and 72\,\% (lower) filled time series to those from an ungapped series 
      of GONG 16$\degr$ tiles.
      The dotted line indicates the value of the duty cycle. The width of the distribution for 72\,\% fill is to be noted, as well as its extension towards high values.
}
\label{fig:Amp_histo_98_72}
\end{figure}

\begin{figure}
\centerline{\includegraphics[width=0.9\columnwidth]{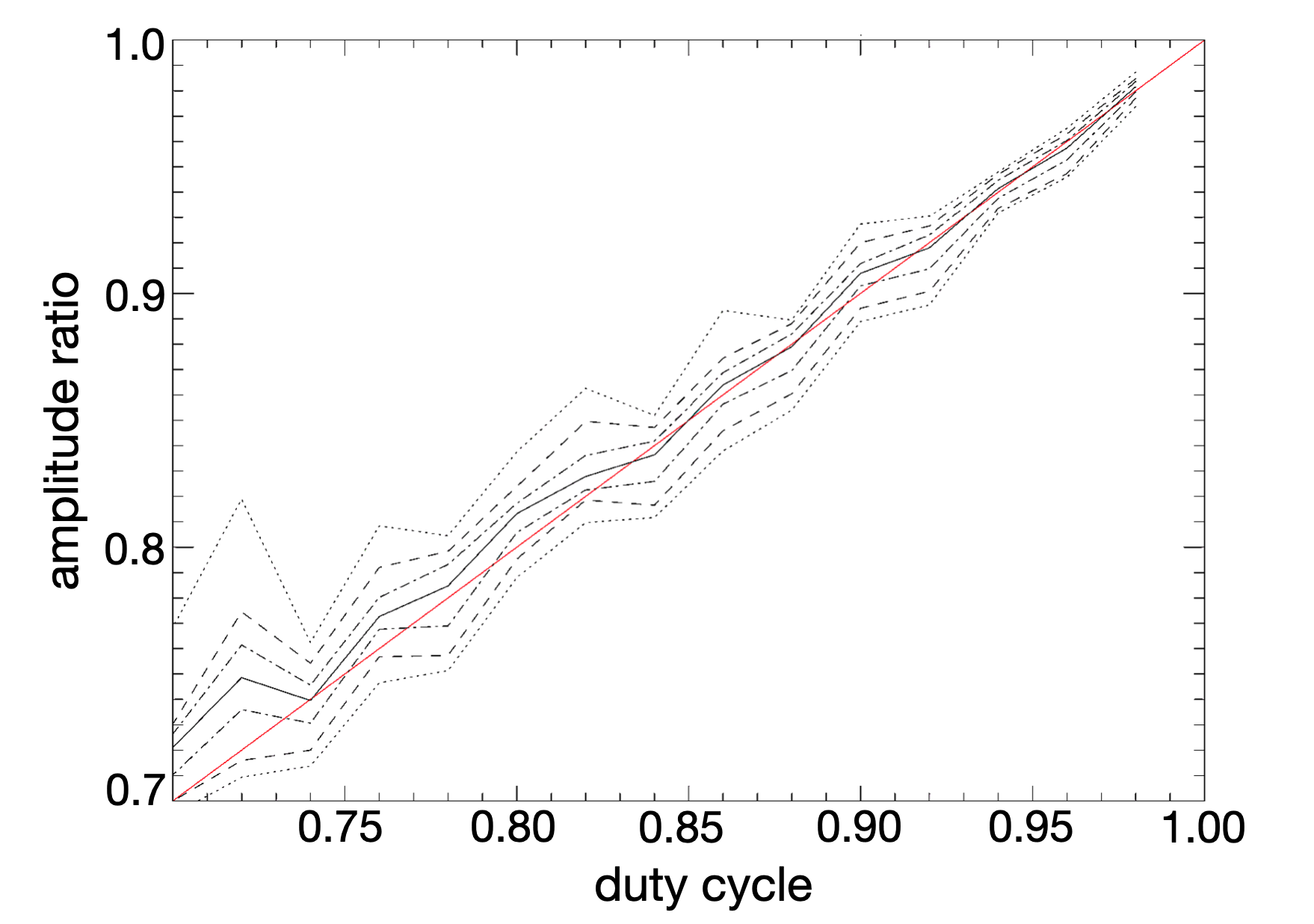}}
\caption{Dispersion of the ratio of fitted amplitudes from
GONG 16-degree
gapped series to those of ungapped series \textit{versus} the duty cycle. The solid line represents the median of fitted amplitudes, and the different dotted and dashed lines 1, 2 and 3$\sigma$. The red line is the expected value (amplitudes varying linearly with the duty cycle).}
\label{fig:percentiles}
\end{figure}

The effect of gaps in the time series is apparently straightforward: the observed
spectrum is simply the convolution of the intrinsic spectrum with the
spectrum of the observing window; mode amplitude will decrease and mode width will increase. 
This effect has been studied by \cite{KommHoweHill_2000} and \cite{BurtsevaTripathyBogart_etal_2013}. An illustration of the correction of this bias is shown in Figure~\ref{fig:mdi_r15_los} for quiet regions observed by MDI, where one can compare its effect in comparison to the line of sight effect.\\
One should not forget that gaps not only decrease the observed amplitude of the signal: dispersion of the measured amplitude is also affected.
We show here that some variance is to be expected when dealing with real cases, in particular when the duty cycle is less than 80\,\%. In order to quantify the effect of the gaps, a 1664-minute long ``ring-day'' of data from the Global Oscillation Network Group (GONG)
with no missing data was analysed after gaps were artificially created (for example using a real window from another ring-day) leading to various values of the duty cycle, from 98\,\% down to 70\,\%.
The top panel of Figure\,\ref{fig:Amp_histo_98_72} shows the distribution of the ratio of the fitted mode amplitudes
from gapped series with a duty cycle of 98\,\% to those from ungapped series. They clearly peak at a value corresponding to the duty cycle, 
with a mild dispersion around this value.
Bottom of Figure\,\ref{fig:Amp_histo_98_72} shows the same ratio for
a duty cycle of 72\,\%. Even if the distribution still peaks at the value of the duty cycle, 
the dispersion is much higher 
in
this particular case, values as high as 1.2 in extreme cases (compared to the expected value of 0.72).
The overall variation of the dispersion for different values of the duty cycle is shown in Figure\,\ref{fig:percentiles}, confirming that fitted amplitudes when the duty cycle is too low have to be taken with caution. This is even more relevant for modes at frequencies where the signal-to-noise ratio (SNR) is low (typically below 2\,mHz or above 4\,mHz).

\section{Observations and results}
\label{sec:results}

\subsection{The archetype case: the ``Halloween'' flare}
\label{sec:Halloween}

One of the most powerful flares observed in the last decades was the so-called Halloween flare. It was the largest solar flare ever recorded by the GOES system.
It is also an archetypal case for the apparent signature of
flares affecting the amplitudes of $p$-modes (of high -- greater than 200 --
spherical harmonic degree [$\ell$])
as reported by \citet{maurya2009}, later extended by
\citet{Maurya_2014}. The former work is focused on the analysis of this `Halloween'
intense flaring active region (October\,--\,November 2003) during Cycle 23. Using ring-diagram analysis of five observing periods 
(ending just before, starting just after and intermediate timings) the most intense flare on 28 October 2003, \citet{maurya2009} shows a clear increase (sometimes more than 100\,\%) of $p$-mode amplitude as fitted in ring diagrams derived from GONG data. However, it should be noted that the highest relative increases are observed at high frequency, where absolute amplitudes are low.
It can also be noted that this flare did not show a sunquake-type
event despite its observed strength (in X-ray flux).\\
\citet{maurya2009} analysed
five data sets with 16$\degr$ in radius and 1664 minutes duration each with different starting times, such that the flare onset 
(SOL2003-10-28T09:51)  
was placed at the very end of Region R1, at three-fourth of R2, centre (R3), one-fourth (R4) and at the beginning (R5). For the five regions
the central meridian distance equal to -18$\degr$ to -3.5$\degr$ (respectively),
with a constant latitude of -22.5$\degr$ while the Sun had a B$_0$ angle of 4.8$\degr$. 
The mode amplitudes, obtained using MDI data, for regions R1\,--\,5 
and for quiet regions at the same disc position (Q1\,--\,5) but at a different time
are shown in Fig.~\ref{fig:mdi_r15_los}.
The l.o.s. variation of these regions, if not corrected, 
could lead to an artificial increase of mode amplitudes,
by as much as 60\,\% as shown in the lower panel for Q1\,--\,5.
The regions have a large MAI: [R1\,--\,5] have MAI equal to 260, 266, 273, 279, and 283\,G, respectively.
The mode amplitude does not change significantly within this MAI range.

We perform here a re-analysis of this event. However, we prefer not to use GONG data since they present duty-cycle values as low as 0.72 during the relevant period of observation, considering the effect of gaps in the data described in Section\,\ref{sec:gaps}. Since Helioseismic and Magnetic Imager (HMI, the latest instrument launched) observations started in 2010, we use here data from the Michelson Doppler Imager (MDI) on board SOHO \citep[HMI predecessor, see][]{scherrer1995} to analyse this strong flare, despite some gaps in the data: the duty cycle is 0.97 for all regions, except for R5 which is 0.81.\\

\begin{figure}
	\includegraphics[width=\columnwidth]{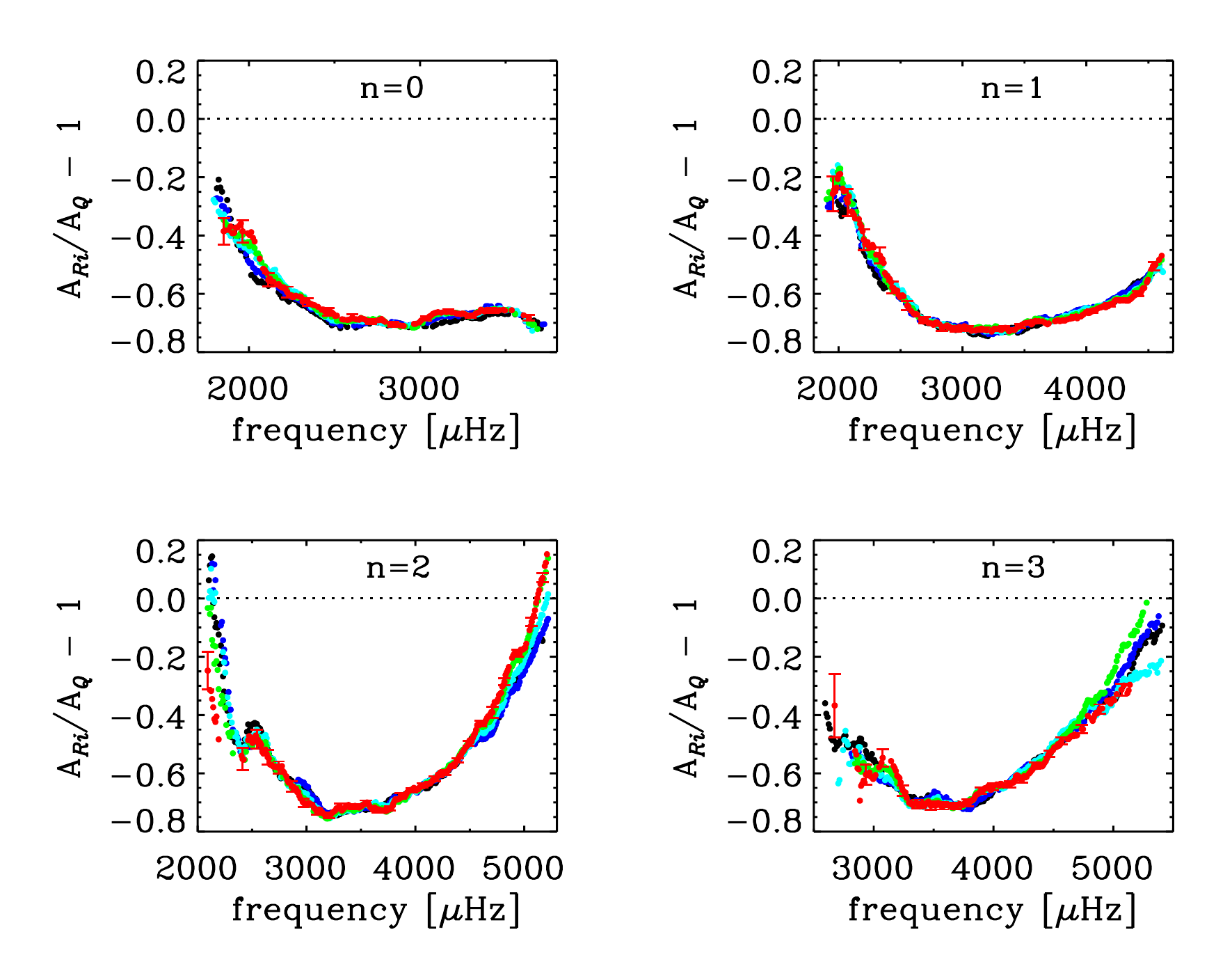}
    \caption{
Relative amplitude of regions R1\,--\,5 in relation to a quiet region
at the corresponding region $R_i$ disk position.
Regions R1\,--\,5 are in black, blue, light blue, green, and red, respectively (using the same colour code as in Figure~\ref{fig:mdi_r15_los}). 
The strong mode reduction due to very large MAI of these regions is clearly seen. 
The coverage correction was applied.
No clear increase of amplitudes from R1 to R5 is seen for these four values of $n$, as well as for $n$=4 and 5 (not shown here).
The error bars are shown only for R5 and for every 10 points.
The errors are smaller than $\sim$0.03, except at small and large frequencies where it can be as large as $\sim$0.1.
}
    \label{fig:halloween-norm}
\end{figure}

We applied the algorithms\footnote{mtrack, pspec3, rdfitc, and maicalc at http://hmi.stanford.edu/teams/rings/modules}
from the HMI ring diagram pipeline \citep{bogart2011a,bogart2011b} to the MDI data\footnote{MDI Full Disk Dopplergram (`mdi.fd$\_$V') available at http://jsoc.stanford.edu/ajax/lookdata.html} at position of the observations on the disc identical to those used by \citet{maurya2009} for the GONG data (Fig.~\ref{fig:mdi_r15_los}).
MDI data have the same cadence 
(1-minute)
and same spatial resolution (128$\times$128 pixels) as GONG data used by \citet{maurya2009}.
Although, it has the same tracking time, MDI image data was mapped using Postel’s azimuthal equidistant projection, while the GONG pipeline uses transverse cylindrical projection. After remapping, the MDI data was tracked at the Carrington rotation rate whereas the GONG pipeline tracks with a differential rotation rate of the Sun \citep{snodgrass1984}.
The power spectra were fitted using the peak-fitting method \citep[called \textit{rdfitc}, see][]{bogart2011a}, which is different than the one used in the GONG pipeline and was chosen because it fits more modes and fits an asymmetric Lorentzian. 
The mode amplitude is the mean power in the ring for a given mode.
It is corrected considering a strict proportionality of amplitudes with the duty cycle (Figure~\ref{fig:percentiles}).
%
The mode amplitudes for the five regions [R1\,--\,5]
are shown in Figure~\ref{fig:mdi_r15_los} as stars.
The influence of the position on the disc is treated by using quiet observations on another date but at the exact same positions (that we call regions Q1–-5). 
We used quiet regions obtained days before the large active region associated with the Halloween flare appeared
and that have MAI$\sim$12\,G, which is considered quiet for a 16$\degr$ tile.
In addition, given that R1-5's MAI varies by more than 20\,G, we did not look for a lower MAI in the 2-3 months each year that the MDI Full Disk Program operated.
Their mode amplitudes are shown in Figure~\ref{fig:mdi_r15_los} (full circles).
Then, the amplitudes fitted during the Halloween sequence (R1\,--\,5) are divided by the amplitudes of the quiet sequence. Results are shown in Figure~\ref{fig:halloween-norm}.

We did not find any significant variation in the mode amplitude before or during the Halloween flare. 
Very different results, a mode amplitude increase by as much as 150\,\%, were found by \citet{maurya2009}, who did correct for the small duty cycle using a method described in \citet{Komm_2000}, but the data were not corrected for line-of-sight variations between Regions 1 to 5.

\subsection{A statistical approach}\label{sec:stat}

\begin{figure}
    \centering
	\includegraphics[width=\columnwidth]{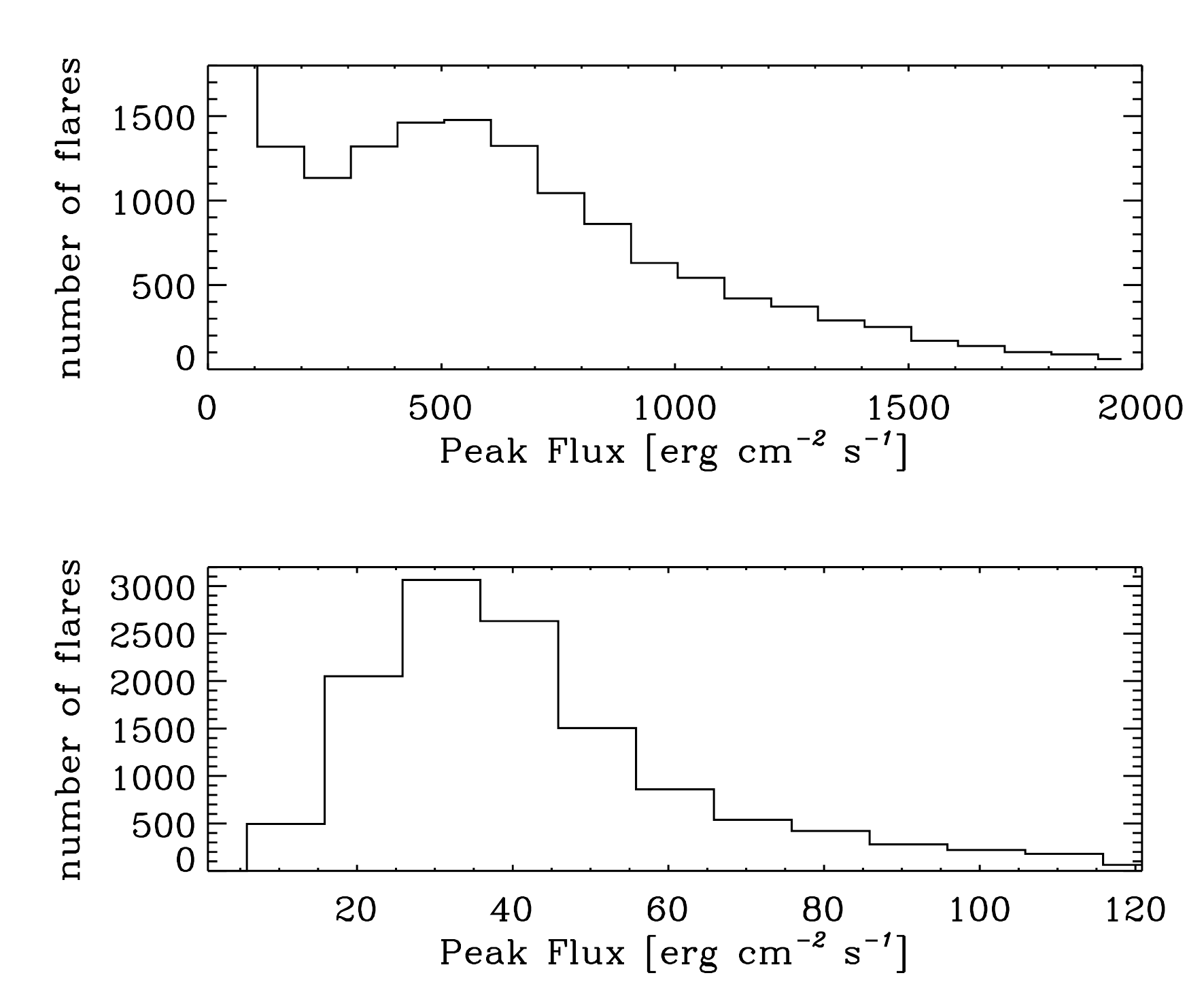}
    \caption{Number of flares in the HEK database detected by \textit{FFT} during four years of highest solar activity during Cycle 24 within $\pm32\degr$ in solar latitude and $\pm32\degr$ in CMD of disc centre.
    The bottom panel shows the large number of flares with small peak flux.
}
    \label{fig:hist_flares}
\end{figure}

Since statistically significant conclusions
cannot be drawn from one event, a large number of events have to be taken into account.
The Heliophysics Event Knowledgebase \citep[HEK, see][]{hurlburt2012}, which is one the largest catalogs of solar events and features \citep{barnes2020}, has several automated solar feature-detection methods applied to solar observations.
We use three automated detection routines on the HEK database: 
\textit{SolarSoft}, \textit{SWPC}, and \textit{Feature Finding Team (FFT)}.
These methods give the flare occurrence times (beginning, end, and peak times) and its location on the solar disc.
The first two return the GOES class \citep[e.g.][]{phillips1995} and the last one gives the peak flux.
\textit{SolarSoft} and \textit{FFT} use data from the 
Atmospheric Imaging Assembly \citep[AIA, see][]{lemen2012}, onboard the Solar Dynamics Observatory (SDO), while \textit{SWPC} uses Geostationary Operational Environmental Satellite (GOES) X-Ray Sensor (XRS) data \citep{garcia1994}.
We analysed flares that occurred during high solar activity during Cycle 24, from June 2012 to August 2016.
Figure\,~\ref{fig:hist_flares} shows the number of flares as a function of the peak flux in the HEK database (provided by \textit{FFT}) that occurred within $\pm 32\degr$ in solar latitude and central meridian distance (CMD) of disc centre, to avoid large line-of-sight effects.

We analysed the acoustic-mode parameters 
obtained using ring-diagram analysis applied to five-degree regions of HMI Dopplergrams \citep{schou2012} and
provided by the HMI pipeline\footnote{`hmi.rdVfitsc$\_$fd05' data series available at http://jsoc.stanford.edu/ajax/lookdata.html} \citep[for details see][]{bogart2011a,bogart2011b}.
The selected regions have 128$\times$128 pixels which allow a circular spatial apodization with a five-degree radius.
Each 
patch is remapped to solar coordinates using Postel's projection and tracked for
767 45-second samples, or 9.5875 hours.
The power spectrum is fitted using the
\cite{basu_antia1999} method (`rdfitc')
to extract the mode parameters.
The analysed five-degree data cubes are equi-spaced by 2.5 degrees in solar latitude and CMD.
The mean magnetic field of each tracked region (MAI)
is calculated in the HMI pipeline by remapping and tracking the correspondent region of the HMI magnetograms and integrating the magnetic flux above the instrumental noise.
We chose to use the five-degree tiles to have a better spatial and temporal resolution when analysing the effect of solar flares. 
Also, note that 
a five-degree HMI region has the same number of pixels as the sixteen-degree MDI region since the resolution of the HMI image is more than three times higher than for MDI, and high-resolution GONG (0.040$\degr$/pixel and 0.125$\degr$/pixel respectively).

The HMI data cubes were divided into three sets:
(i) `Strong Flares': those with peak flux larger than 1200 erg\,cm$^{-2}$~s$^{-1}$ and/or GOES class M or X,
(ii) `Weak Flares': with peak flux between 10 and 60 erg\,cm$^{-2}$~s$^{-1}$, and (iii) `Without Flares': where no flare was detected. 
Table~\ref{tab:data_sets} 
shows the number of flares detected, the number of HMI data cubes, and the number of cubes that were used in our analysis.
The same solar flare could have been detected and be present more than once in the HEK database. 
The lower limit of the Strong Flare set was set to 1200 erg\,cm$^{-2}$~s$^{-1}$ in order to have a large enough number of events for a statistical analysis. As a comparison to the analysis of the effect of these more energetic flares on mode amplitudes, we built a Weak Flares set corresponding to much smaller peak-flux again to get a sufficiently large number of events for a statistical analysis (Fig.\,\ref{fig:hist_flares} bottom panel).
When building the Strong and Weak Flare subsets, we considered only data cubes where the flare occurred within 80\,\% of its radius and within the central 90\,\% of its duration.

\begin{figure}
    \centering
	\includegraphics[width=\columnwidth]{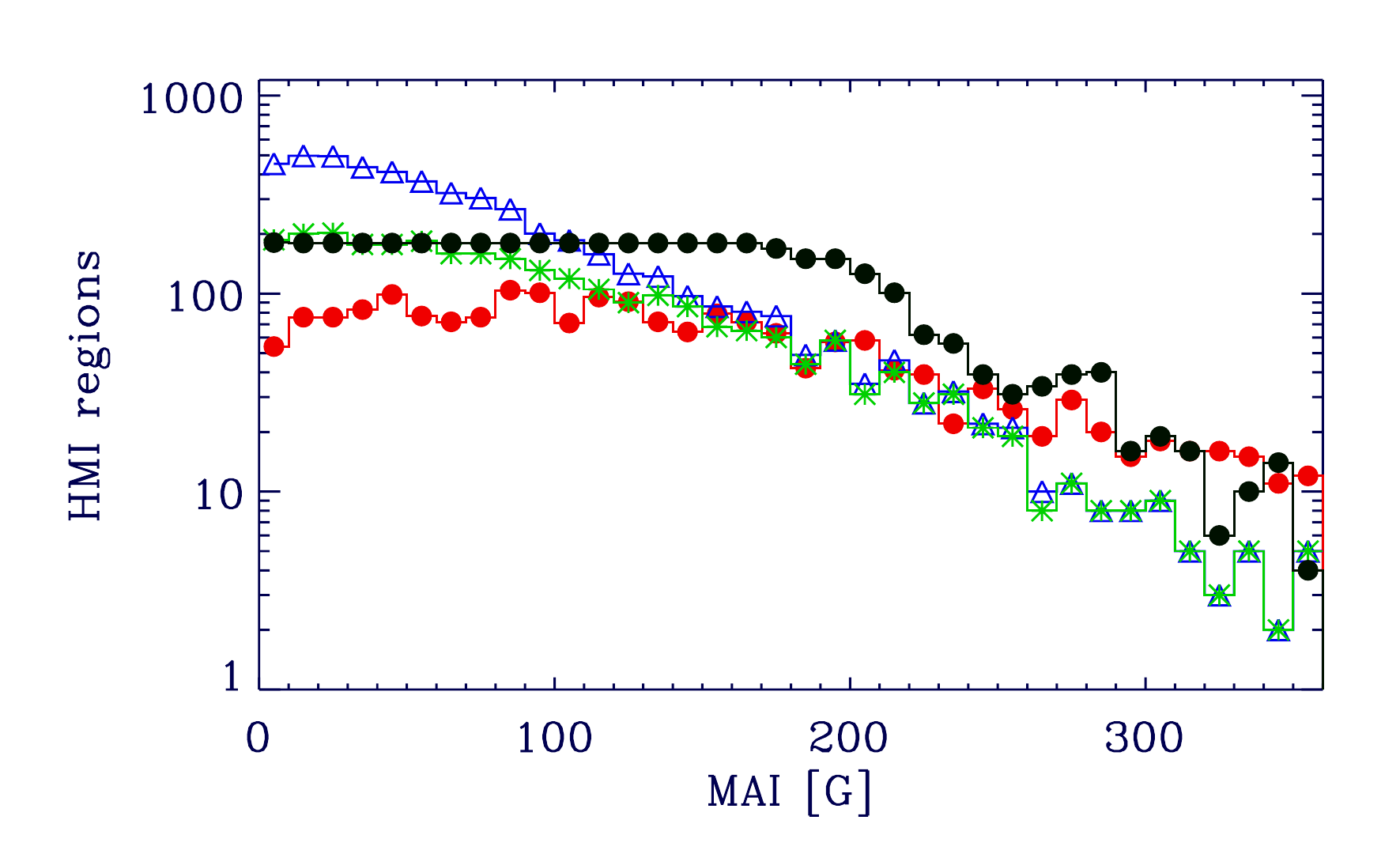}
    \caption{
    Number of HMI data cubes for each data set as a function of MAI. The Without Flares in black circles, Weak Flares in green stars and Strong Flares in red circles, used in our analysis.
    The total number for the Weak Flare set is shown as blue triangles.
    A smaller number of HMI regions for the Weak Flare and Without Flare sets were selected,
    with MAI values evenly distributed, to a number closer to the Strong Flare set.
        }
    \label{fig:plot_mai}
\end{figure}

\begin{figure}
    \centering
	\includegraphics[width=\columnwidth]{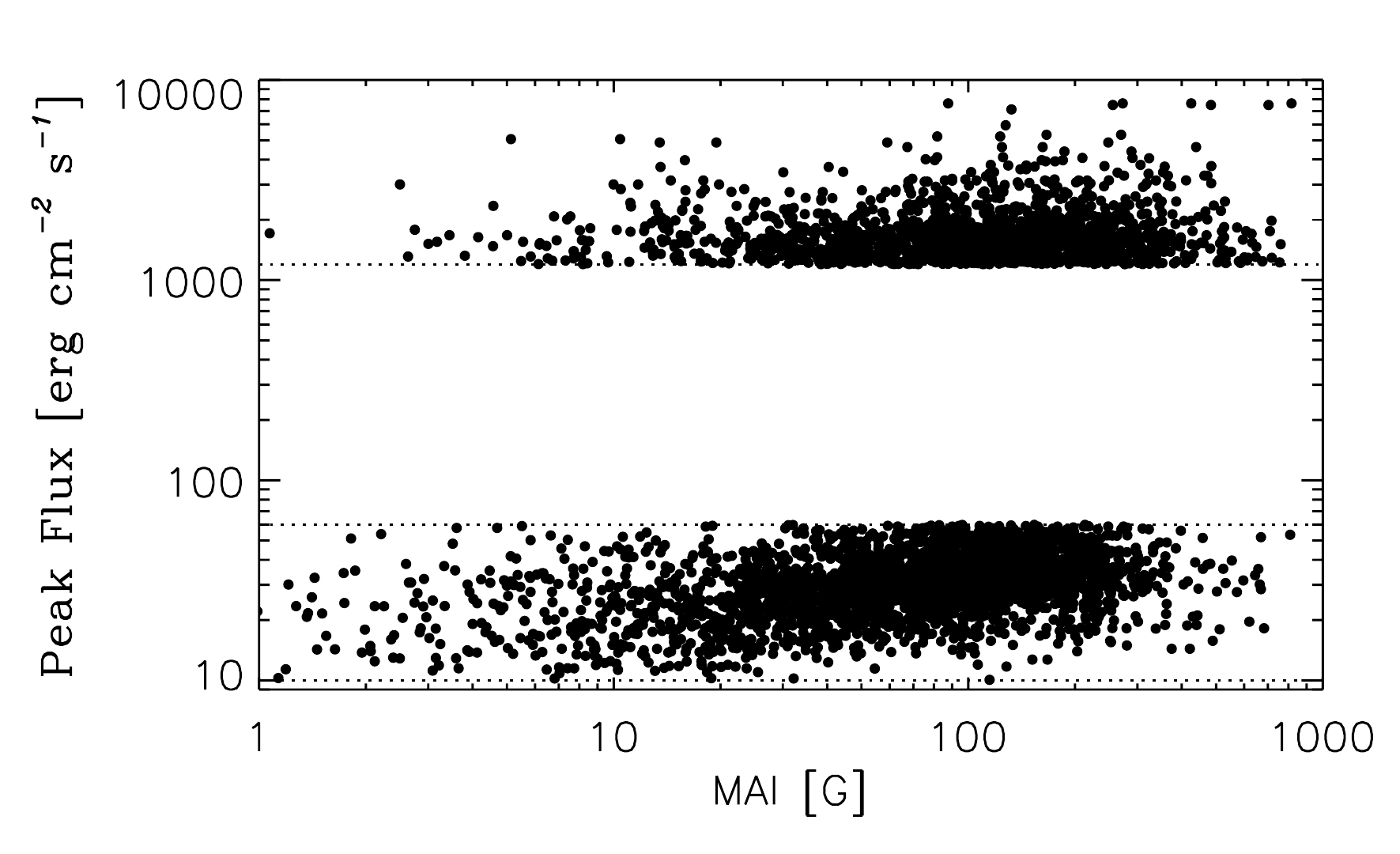}
    \caption{Flare peak flux as a function of the MAI for the Strong and Weak Flare set.
        }
    \label{fig:plot_mai_flux}
\end{figure}

In the case of the Weak Flares, only regions where there are no flares stronger than
60 erg\,cm$^{-2}$~s$^{-1}$ within 1.5 times the region's radius from its centre were considered.
The Without Flare set includes only data cubes without any flares detected by HEK within 1.5 times the tile's radius and within 1.2 times the duration of the data cube.
Figure\,\ref{fig:plot_mai} shows the distribution of MAI values for each data set in Table~\ref{tab:data_sets} at 10\,G intervals.
Most of the Strong Flares regions (75\,\%) have an MAI smaller than 200\,G and only 10\,\% larger than 290\,G.
However, for MAI\,>\,240\,G, there are more regions with Strong Flares than regions with Weak Flares, and for MAI > 300\,G their number is similar to that of regions Without Flares.
Although there are fewer regions of Strong Flares than Weak Flares for MAI\,<\,180\,G,
there are still a large number with Strong Flares with MAI as small as 1\,G.
The fact that flares occur in low MAI region could be surprising. We can however speculate that the complex geometry (flux tube inclination, distant loop footpoints, ...) of magnetic structures at the origin of the flares can explain the low value of the MAI.
We selected a smaller number of HMI regions with MAI values evenly distributed and with high data coverage,
for Weak Flare and Without Flare sets due to their large number. Because of systematic effects, using a larger number would not improve dispersion.
Figure\,\ref{fig:plot_mai_flux} presents the flare peak flux as a function of the MAI of the region where it was detected.
Most flares are in tiles with MAI = 26$-$103\,G and 64$-$190\,G for the Weak and Strong Flares set, 
given by the first and third quartiles.
We do not find any correlation between flare peak flux and 
the MAI of the corresponding HMI tile for the Strong Flare set. For tiles with MAI larger than $\sim$20\,G, the smallest peak flux increases with MAI for the Weak Flare set.

\begin{table}
	\centering
	\caption{Three sets of HMI data analysed.}
	\label{tab:data_sets}
	\begin{tabular}{lrrcr} 
		\hline
		Data set & Flares & HMI regions & Selected & Data Coverage\\
		\hline
		Without Flares & 0 & 11886 & 4143 & 1.00\\   
		Weak Flares & 9905 & 5081 & 2790 & 1.00\\       
		Strong Flares & 2117 & 2007 & 2007 & 0.98\,--\,1.00\\    
		\hline
	\end{tabular}
\end{table}

As discussed in Section\,\ref{sec:line-of-sight}, even a small change in the disc position could be erroneously interpreted as an effect of a flare. Thus, for each selected region in the three sets (Table~\ref{tab:data_sets}), the line-of-sight effect on the mode amplitude due to the observed position on the solar disc has to be removed.
In order to make this correction, we divided the amplitude of each mode ($n$, $\nu$) of each region $i$ by an average of the quiet mode amplitude for the same region's disc position (given by its solar latitude and CMD):
\begin{equation}
\Delta\!A\!/\!A_{_Q} (n,\nu,\text{MAI}) = \frac{A_i(n,\nu, \text{latitude}, \text{CMD}, \text{MAI})}{\overline{A}_Q (n, \nu, \text{latitude}, \text{CMD})} \, - \, 1
\end{equation}
where $n$ is the mode radial order and $\nu$ its frequency.
We used twenty magnetically quiet regions (MAI < 1\,G),
with no flare activity detected, at each solar disc position with 100\,\% data coverage to calculate the quiet mode amplitude average $\overline{A}_Q$ (Figure\,\ref{fig:freq_bias_cris}).
As HMI has been continuously observing the solar full disc since 2010 not only with high spatial resolution but also high data coverage, we were able to impose a very low limit on the MAI of the quiet regions.

\begin{figure}
    \centering
	\includegraphics[width=0.85\columnwidth]{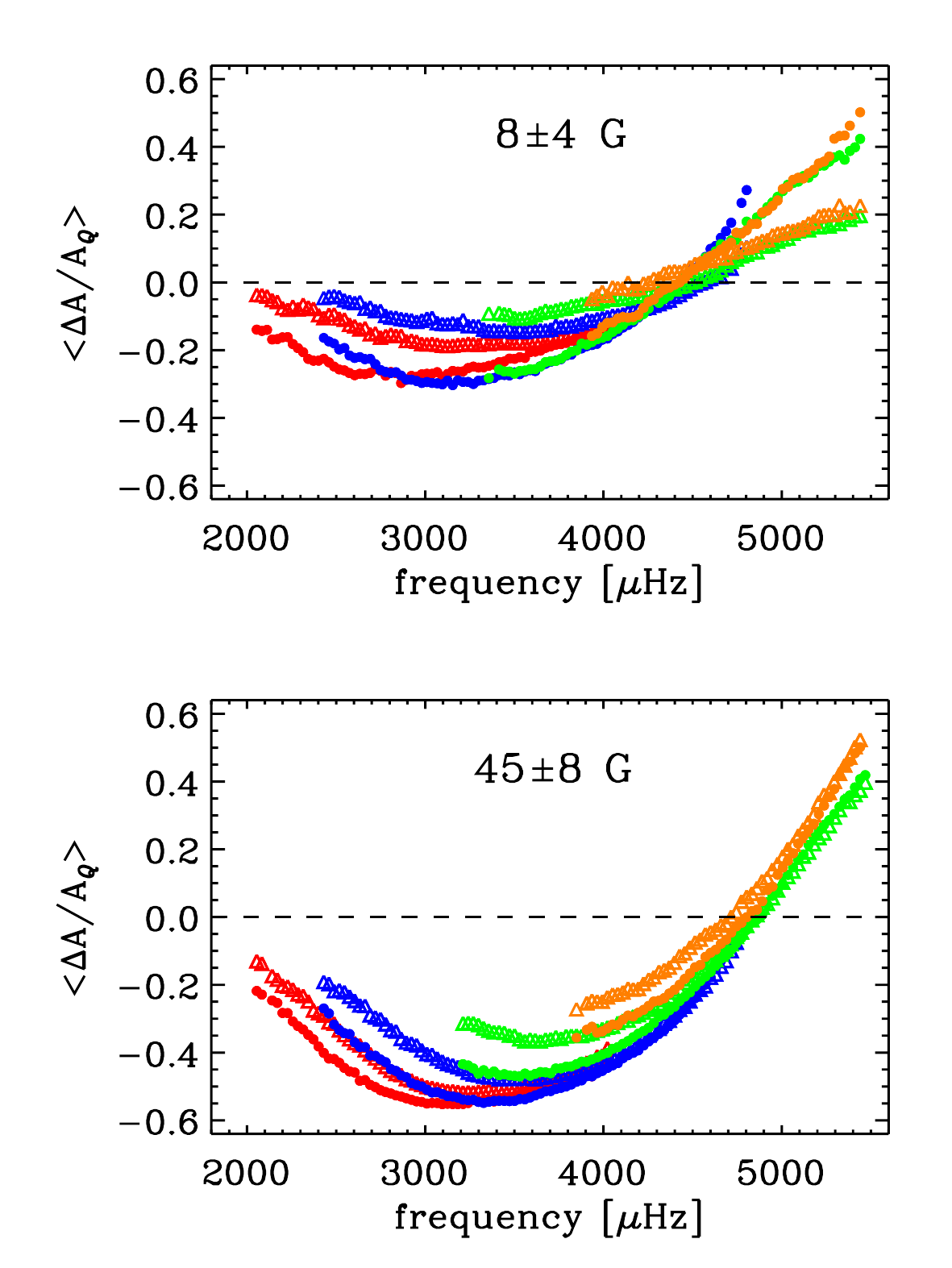}
    \caption{Relative amplitude averaged over HMI regions with MAI in two intervals: (8$\pm$4)\,G (top panel) and (45$\pm$8)\,G (bottom), for Strong Flares (full circles) and Without Flares (triangles) regions. The different colours represent different mode order $n$: 0 (red), 1 (blue), 2 (green), 3 (orange). The error of the mean is of the order of 0.01 and smaller.}
    \label{fig:ampli_freq}
\end{figure}

To look for an effect of a flare on the mode amplitude, we compared the relative mode amplitude, 
$\Delta\!A\!/\!A_{_Q} (n,\nu,\text{MAI})$,
obtained in the Strong and Weak Flares set with those in the Without Flare set.
However, to take into account the effect of strong magnetic fields on the mode amplitude (discussed in Section\,\ref{sec:mai}), we need to compare regions with similar MAI.
Figure~\ref{fig:ampli_freq} shows the average relative amplitude for two MAI intervals: (8$\pm$4)\,G and (45$\pm$8)\,G for regions
Without Flares and regions with Strong Flares. 
The well-known 
decrease of mode amplitude with an increase in MAI
is clearly seen when comparing the top and bottom panels.
In both cases, there is a clear difference between regions with and without flares.
The flaring regions have a smaller amplitude for modes with frequency smaller than $\sim4500\mu$Hz and the opposite behaviour at higher frequencies.

\begin{figure*}
\includegraphics[width=\textwidth]{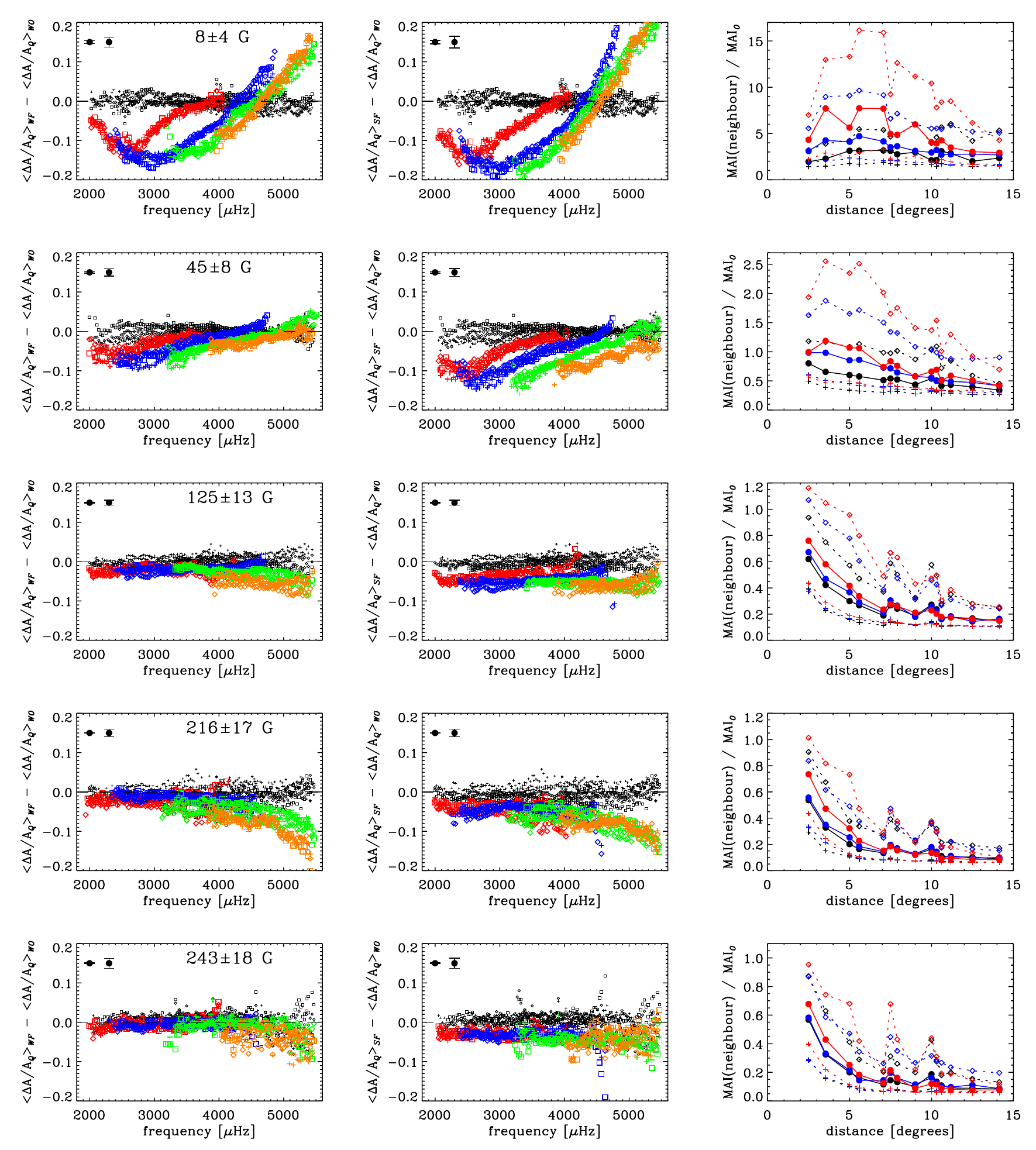}
    \caption{Mean relative amplitude difference between Weak ("WF": left column) and Strong ("SF": centre column) compared to Without Flares ("WO") set. Each row shows the difference for distinct MAI intervals: MAI$_0 \pm \Delta$MAI = 8$\pm$4\,G, 45$\pm$8\,G, 125$\pm$13\,G, 216$\pm$17\,G, and 243$\pm$18\,G from top to bottom. 
    The different colours represent different mode order $n$: 0 (red), 1 (blue), 2 (green), and 3 (orange). 
    The Without Flare set was divided into three sets
    and the mean relative amplitude difference between them was calculated as an indication of systematic variation in our analysis (crosses, diamonds and squares).
    At the top left of each plot is shown the smallest and largest error bars.
    Right column: The median of the MAI in the neighbourhood of the analysed regions,
    divided by the MAI of the target, MAI$_0$, 
    as a function of distance from the target represented by full circles: Without Flares in black, Strong Flares in red, and Weak Flares set in blue. 
    The small crosses and diamonds are the first and third quartiles, respectively.
    }
    \label{fig:delampli_freq}
\end{figure*}

The left and centre columns in Figure~\ref{fig:delampli_freq} show the difference between the mean relative amplitude for flaring and non-flaring regions at five different MAI intervals. 
The mean relative amplitude for each MAI interval, <$\Delta\!A\!/\!A_{_Q}$>,
was calculated using between 30 to 220 regions.
The smallest number is in the bottom panel, since
there are fewer HMI regions with MAI larger than $\sim$220\,G (Figure~\ref{fig:plot_mai}).
The first two lines in the centre column correspond to the MAI intervals shown in Figure~\ref{fig:ampli_freq}.
To estimate systematic and random errors in our results, the Without Flares set was divided into three separate sets.
The Without Flares set is classified according to the value of the
MAI alternating each region between subsets so that each one has a similar number of regions in any MAI range.
The mean mode amplitude difference between them is shown as small black symbols in  Figure~\ref{fig:delampli_freq} for the five MAI intervals.
In Figure~\ref{fig:delampli_freq}, the differences of relative amplitude should be small since these amplitudes are observed in regions with similar MAI. However this is not the case, especially for regions with low MAI.
In regions with Strong Flares (centre column), the mode absorption is even larger than for Weak Flares (left column).
A possible explanation for the mode amplitude decrease observed in these flaring regions 
could be due to the strong magnetic fields usually present in and around a flare.
\citet{rabellosoares2016} calculated mode amplitude variation for five-degree quiet regions when there was an active region (with MAI\,>\,100\,G) in their vicinity, at eight degrees or less centre-to-centre, in comparison with quiet regions at the same solar disc location for which there were no neighbouring active regions (i.e., no region with MAI larger than 5\,G).
They found a mode reduction as large as 0.1 around 3000 $\mu$Hz followed by an enhancement at high frequencies of $\sim$0.15 for $\nu > 5000\,\mu$Hz (as shown in the top left panel in their Figure\,1).
Their results can be directly compared with ours in Fig.\,\ref{fig:delampli_freq}.
Looking for a similar effect in our data, we show
the median MAI in the neighbourhood (within 14$\degr$) of the regions analysed as a function of distance to its centre (rightmost column in Figure~\ref{fig:delampli_freq}).
As one might expect, the vicinity of flaring regions is typically very magnetically active and has large values of MAI, more so for the Strong Flare set. 
The smaller the MAI, the larger is the magnetic field in their surroundings (right column), and the larger the mean relative amplitude difference (left and centre columns).

\begin{figure}
    \centering
	    \includegraphics
	    [width=0.85\columnwidth]{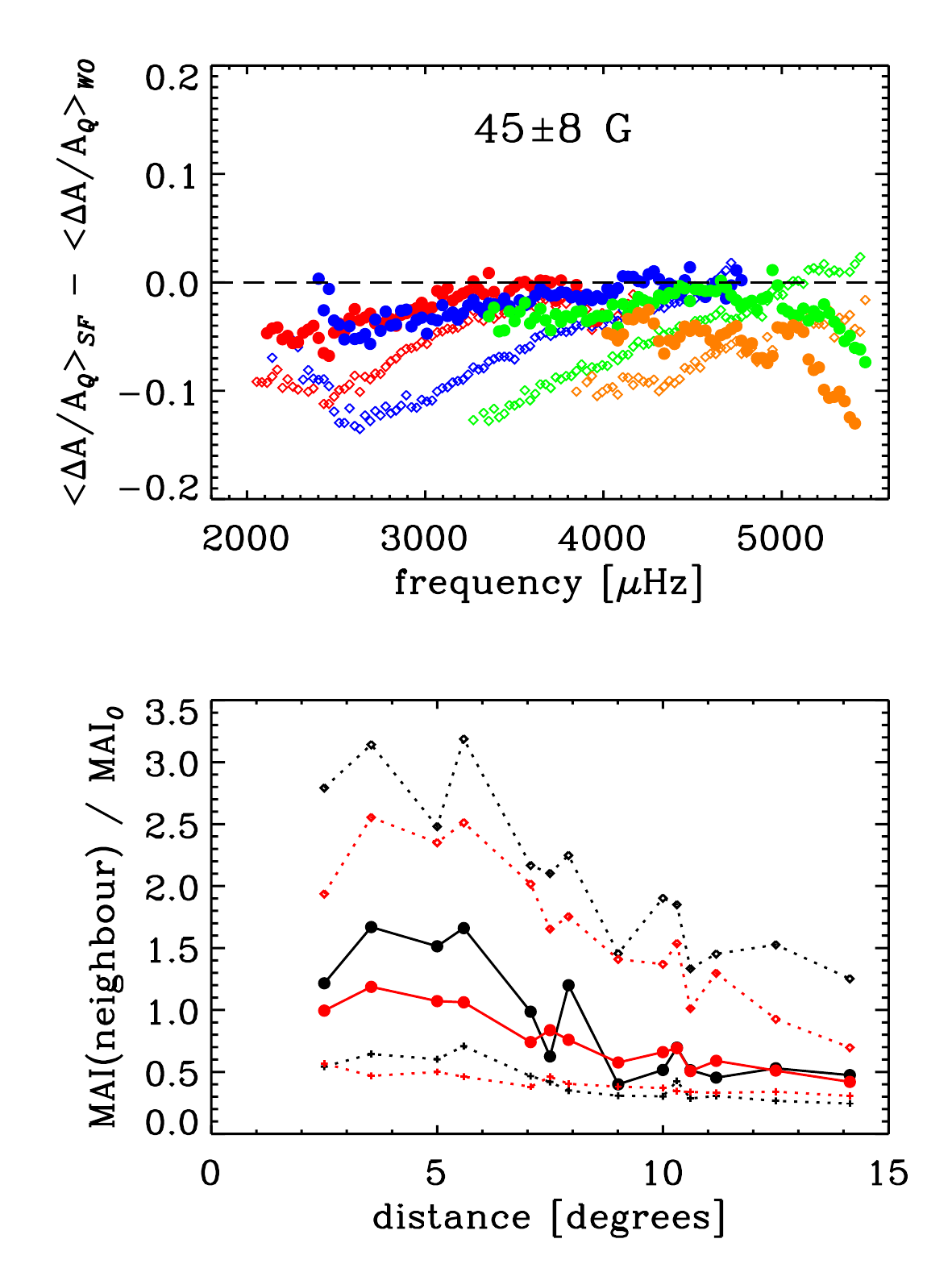}
    \caption{
    Top Panel: Mean relative amplitude difference between Strong Flares (SF) and Without Flares (WO) sets for regions with MAI = 45$\pm$8\,G. The diamonds are using all of the Without Flares set regions in the MAI range, as in Figure~\ref{fig:delampli_freq} (middle column, second line) and the full circles are using only those that have the mean MAI in their vicinity 
    in the 90\textsuperscript{th} percentile (see text).
    The different colours represent different mode order $n$: 0 (red), 1 (blue), 2 (green), and 3 (orange).
    Bottom panel: The median, first, and third quartiles of the MAI in the neighbourhood of the analysed regions divided by the MAI of the target (i.e. 45\,G) as a function of the distance shown as full circles, crosses, and diamonds respectively. Those for the Strong Flares regions are in red and are the same as in Figure~\ref{fig:delampli_freq}. Those for the selected regions in the Without Flares set are in black.
    }
    \label{fig:comp_strong}
\end{figure}

To check the hypothesis that the nearby active regions are affecting the mean relative amplitude differences, we selected the regions in the Without Flares set
with the largest mean MAI in their vicinity, those in the 90\textsuperscript{th} percentile, and recalculated the mean relative amplitude difference. Figure~\ref{fig:comp_strong} compares the difference shown in Figure~\ref{fig:delampli_freq} (diamonds) with the new difference (full circles) which are considerably smaller in absolute value, confirming the effect of the surroundings on our results. The medians MAI in their neighbourhood are now very similar to each other (bottom panel).

\begin{figure}
    \centering
	    \includegraphics[width=\columnwidth]{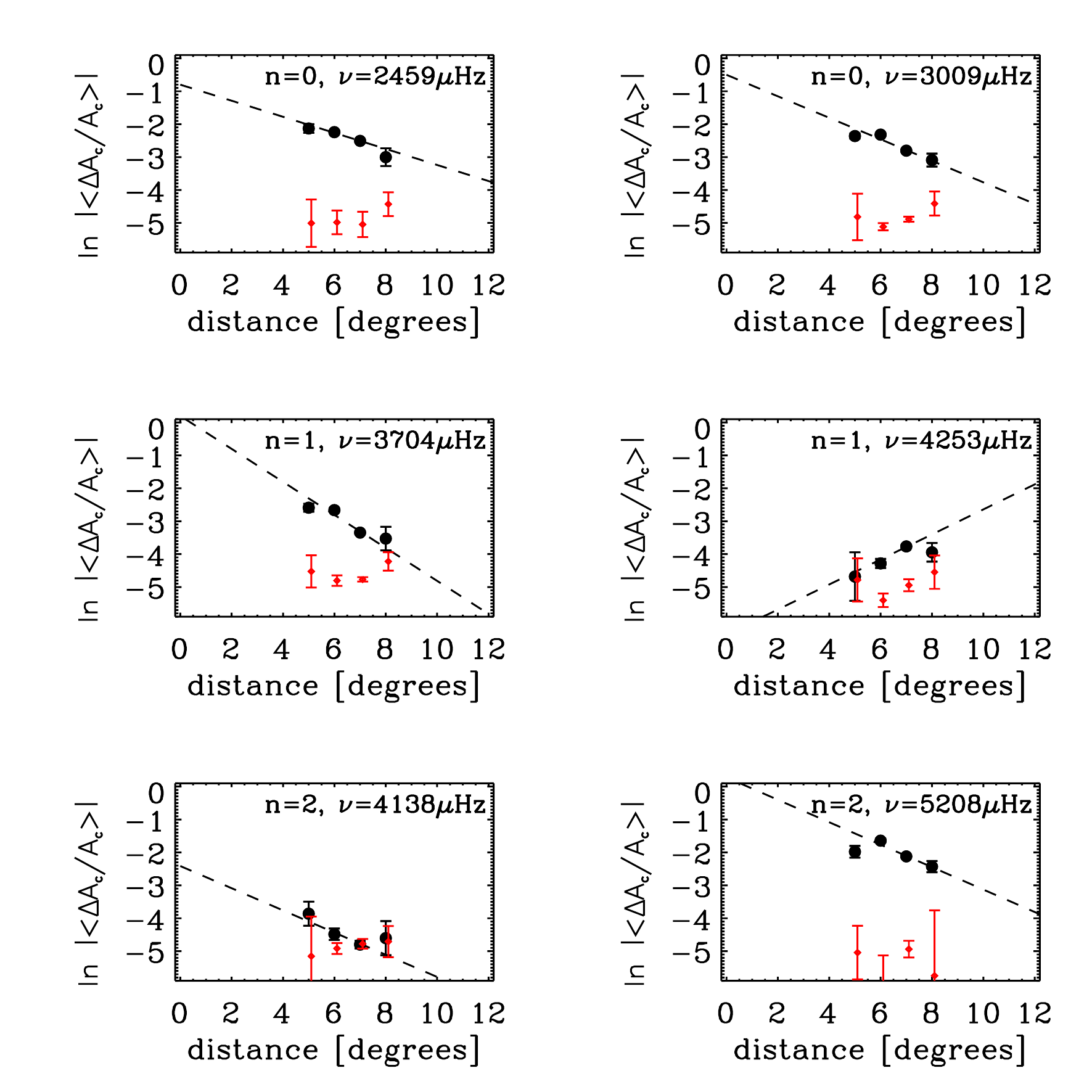}
	        \caption{
Examples of the logarithm of relative amplitude variation absolute value with distance for six different modes \citep[adapted from][]{rabellosoares2018}. 
Top panels: $n$=0 and $\nu$=2460 (left) and 3009\,$\mu$Hz (right). 
Middle panels: $n$=1 and $\nu$=3704 and 4253\,$\mu$Hz.
Bottom panels: $n$=2 and $\nu$=4138 and 5208\,$\mu$Hz.
The dashed line shows the linear fit. 
The small red symbols show the results for a control set \citep[described in][]{rabellosoares2018},
that gives an indication of the noise level of the analysis.
    }
    \label{fig:apj18a}
\end{figure}

\begin{figure}
    \centering
	    \includegraphics[width=0.95\columnwidth]{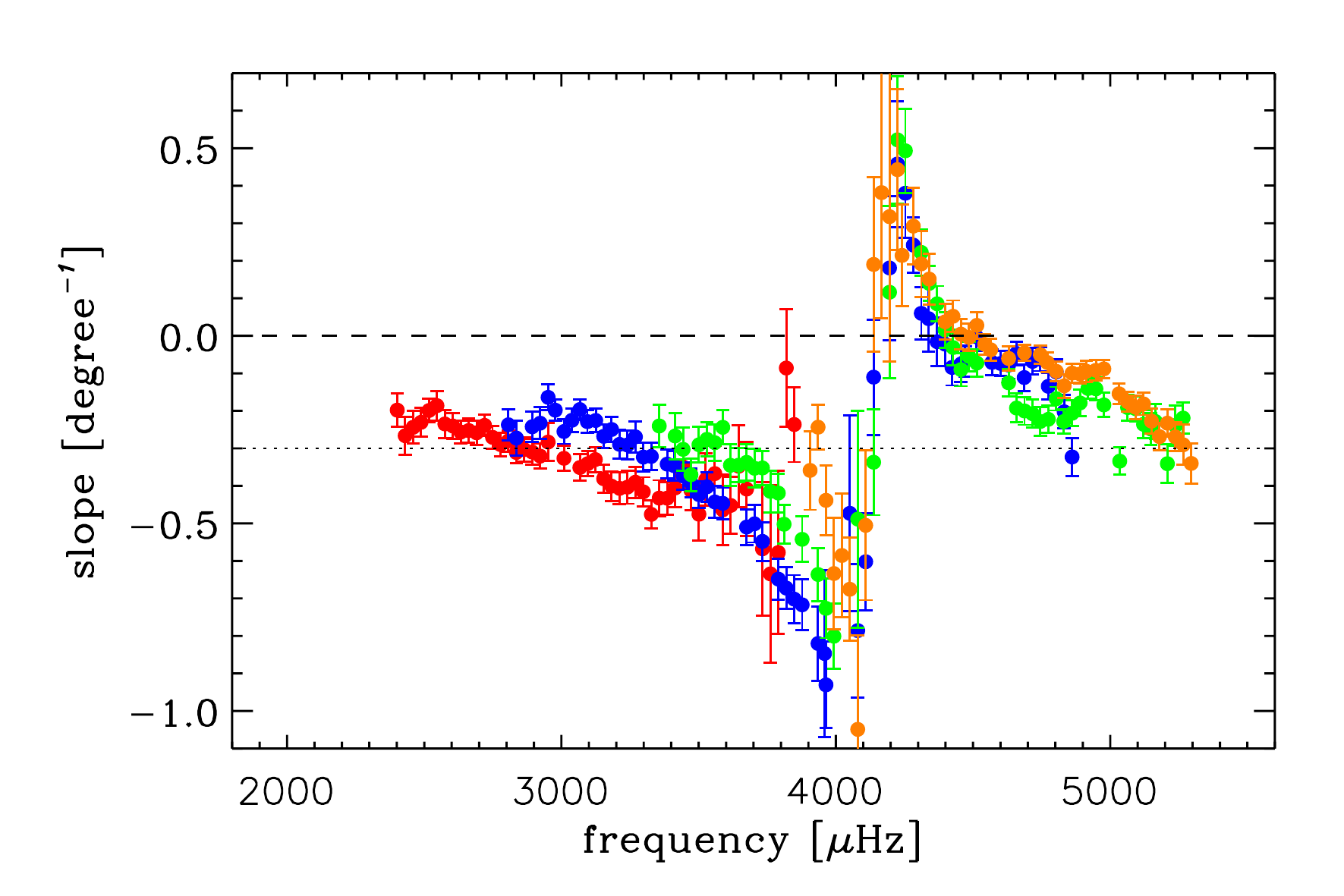}
    \caption{
The slope of the linear fit as a function of mode frequency
\citep{rabellosoares2018}. Modes with $n$=0, 1, 2, and 3 are in red, blue, green, and orange respectively.  The large variation seen in the linear-fit parameters around approximately 4100\,--\,4500 $\mu$Hz corresponds to relative amplitudes very close to zero 
shown in Figure~\ref{fig:apj18a} \citep[and, in more detail, Fig. 1 in][]{rabellosoares2016}. Hence the fitting is not meaningful.
The horizontal dotted line corresponds to the mean and is equal to -0.3 degree$^{-1}$.
}
    \label{fig:apj18b}
\end{figure}

To take into account this additional influence on the mode amplitude, for each region in the Weak and Strong Flares sets, we looked for a region in the Without Flare set 
with not only similar MAI but also with a similar MAI in its vicinity.
Since the effect of a nearby active region varies with its distance to the analysed region, 
it is necessary to quantify this effect. This was done by
\cite{rabellosoares2018} that extended their previous work \citep{rabellosoares2016}
and 
analysed
how the effect of an active region on a quiet region varied with their separation.
Figure~\ref{fig:apj18a}, adapted from Figure~2 in \cite{rabellosoares2018}, shows the variation for six different modes
and
Figure~\ref{fig:apj18b} shows the slope of a linear fit for each mode.
The decrease in the influence of the nearby active region with its distance, given by the exponential coefficient (Figure~\ref{fig:apj18b}), varies for each mode.
We adopted an average value of -0.3 for the exponential coefficient.\\
Then, to calculate the relative amplitude difference between a flaring and a non-flaring region,
we chose a non-flaring region with a neighbourhood most similar to each region in the Weak and Strong Flares set, as the one that minimises the function:
\begin{equation}
      \chi^2 = \frac{\sum_i \left\{ \left[ \text{MAI}_{\text{flaring}}(d_i) - \text{MAI}_{\text{non-flaring}}(d_i) \right] \times \mathrm{e}^{-0.3 d_i} \right\}^2}
   {\sum_i \left[e^{-0.3 d_i} \right]^2}
   \label{eq:chi2}
\end{equation}
where $d_i$ is the distance of the nearby active region
and varies from 0 to 8 degrees.\\

\begin{figure}
   \centering
    \includegraphics[width=0.8\columnwidth]{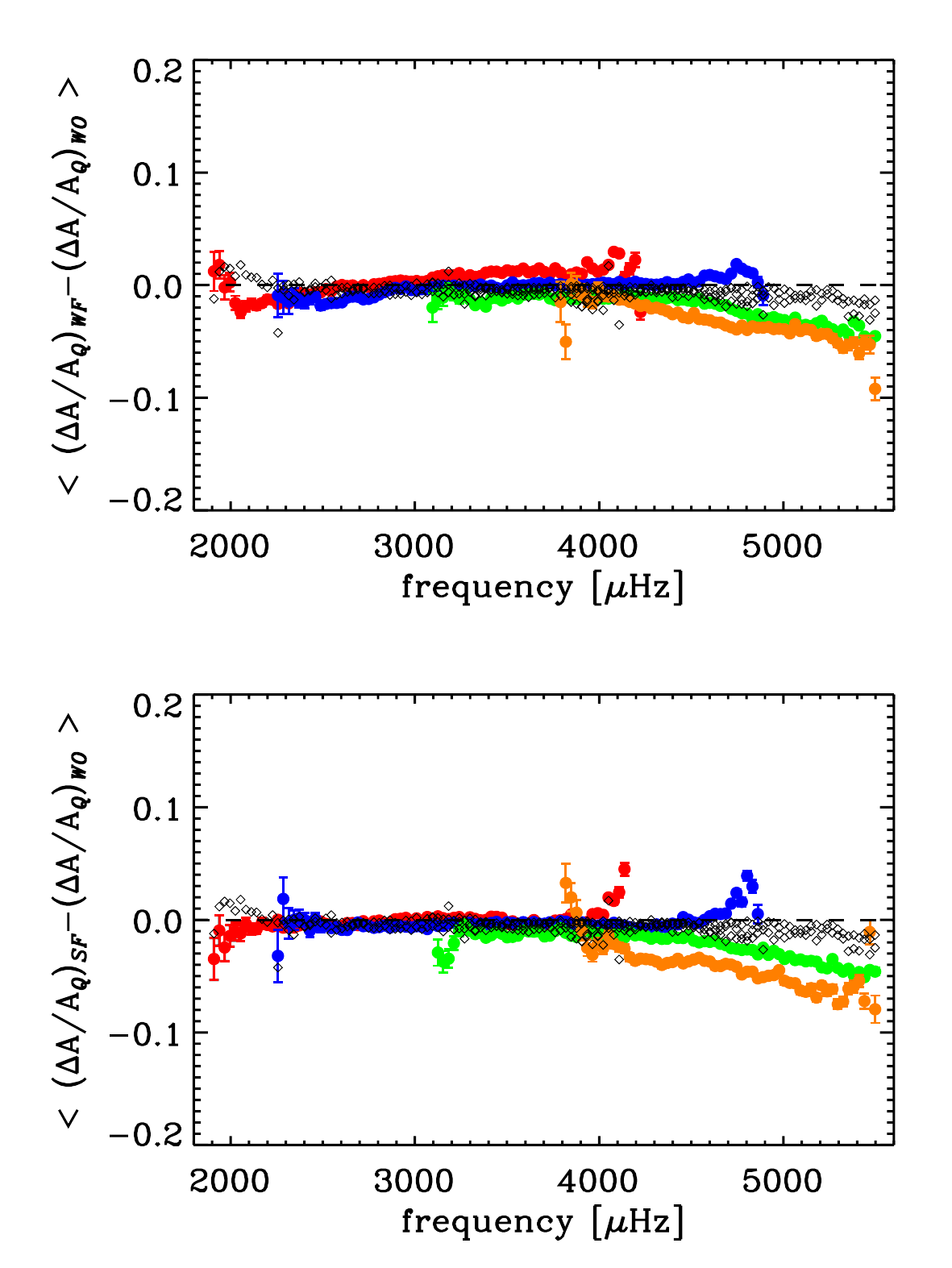}
    \caption{Mean relative-amplitude difference between all pairs of flaring and non-flaring regions (WO) with similar MAI in and around each region. The Weak Flares (WF) set is shown in the 
    top panel and the Strong Flares (SF) on the 
    bottom.
    Modes with $n$ from zero to three are given by red, blue, green, and orange symbols. The error bars are the error of the mean for each mode. 
    The mean relative-amplitude difference between pairs of two distinct groups of the Without Flare set is shown as black diamonds.
    }
    \label{fig:pairs}
\end{figure}

\begin{figure}
   \centering
    \includegraphics[width=0.9\columnwidth]{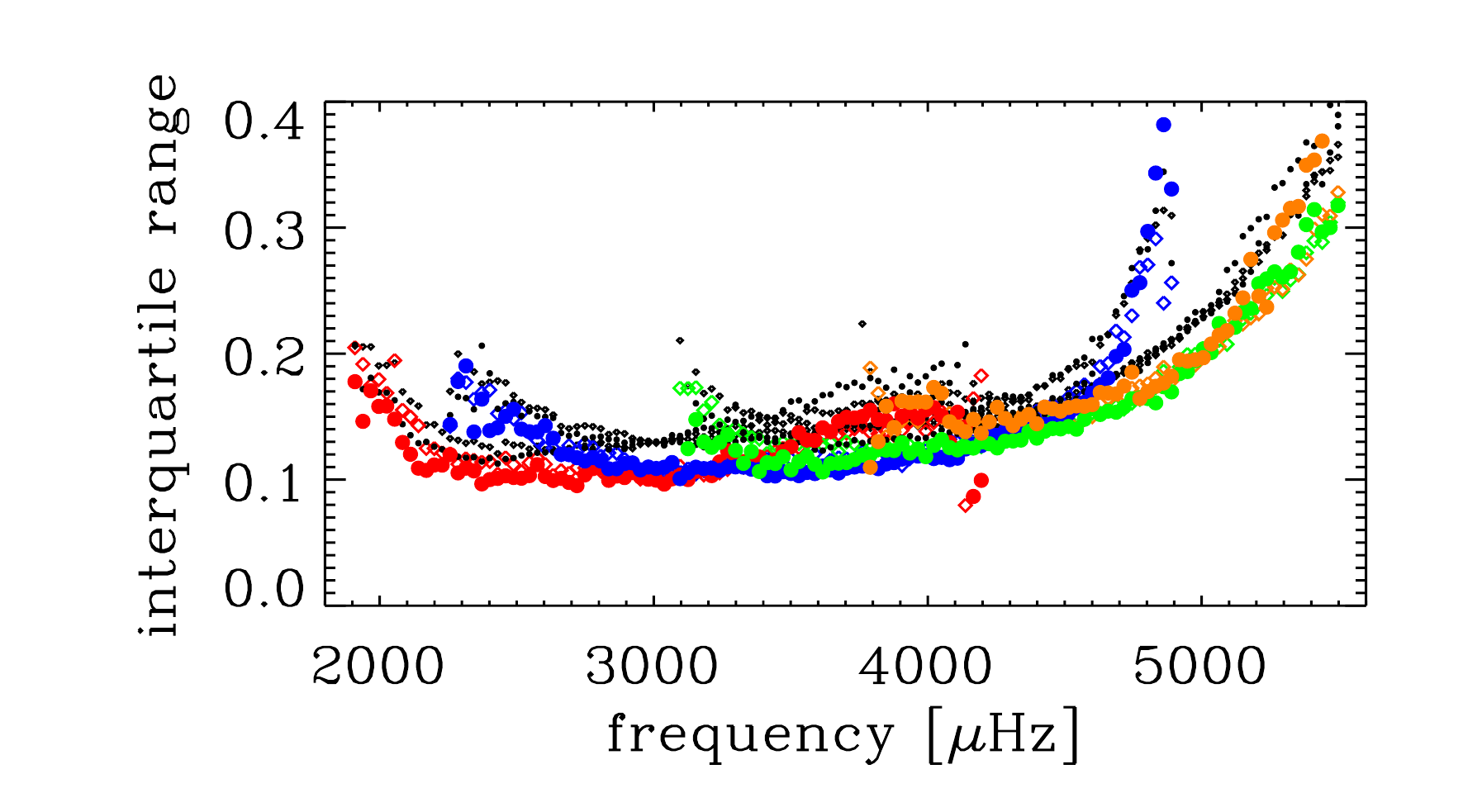}
    \caption{
    The interquartile range
    of the relative-amplitude pair difference:
    ($\Delta\!A\!/\!A_{_Q}$)$_{\text{WF or SF}}$ - ($\Delta\!A\!/\!A_{_Q}$)$_{\text{WO}}$
    for each mode,
    where the Weak Flares are given by diamonds and the Strong Flares by full circles. 
    The different colours are for the different order $n$.}
    \label{fig:quartiles}
\end{figure}

Taking into account all the corrections listed above, it becomes possible to reliably estimate the potential impact of flares on the amplitude of acoustic modes. In Fig.\,\ref{fig:pairs}, we show the average of the relative-amplitude difference for each mode between all pairs with $\chi^2$ in the 40\textsuperscript{th} percentile, which corresponds to $\sqrt{\chi^2} \leq$ 12\,G and 16\,G for the Weak and Strong Flares set, respectively.
The difference in the maximum $\chi^2$ between Strong and Weak Flare sets indicates
that the vicinity of the Without Flare set is, in general, different from the Strong Flare set
(as shown in the right column in Figure~\ref{fig:delampli_freq}).
For each mode, in Fig.\,\ref{fig:pairs}, the number of averaged regions in the Weak (Strong) Flares set varies from 25 to 834 (554)
and the median number of regions is equal to 660 (405).
We chose the 40\textsuperscript{th} percentile in order to have at least 20 regions for each mode and a $\sqrt{\chi^2}$ not much larger than the adopted MAI intervals at the analysed region (shown in Fig.\,\ref{fig:delampli_freq}). The results in Fig.\,\ref{fig:pairs} are similar to those obtained with a smaller percentile.
Once again, to estimate errors in our results, 
we divided the Without Flares set into two separate groups and
calculated the average relative-amplitude difference between them
for $\sqrt{\chi^2} \leq$ 16\,G (shown as small black symbols in Figure~\ref{fig:pairs}).
Our results show an average close to zero for both sets (Weak and Strong Flares)
indicating that our minimisation (Equation~\ref{eq:chi2}) is a reasonable approximation.
Using a different exponential coefficient 
to describe the effect of neighbour magnetic regions on observed amplitudes, -0.2 or -0.5, does not change the results 
appreciably.
Figure~\ref{fig:quartiles} indicates, for each mode, the dispersion of the relative-amplitude difference of the pairs of regions analysed.
For each mode, the skewness of the relative-amplitude difference is close to zero. A negative skewness (i.e., a median larger than the average) would imply that more than half of the regions have an  positive relative-amplitude difference, indicating an increase in the mode amplitude during a flare. The kurtosis is slightly positive, suggesting a small tail in the distribution.
As shown in Fig.~\ref{fig:quartiles}, there is a sharp increase in the dispersion of the relative-amplitude difference for modes with frequency larger than ~4700\,$\mu$Hz. The acoustic-mode amplitudes are an order of magnitude smaller (or even less) for these high-frequency modes (see Fig.~\ref{fig:mdi_r15_los}). Another issue that could contribute to the increase in variability is that \citet{rabellosoares2016} observed that the mode amplification due to nearby active regions starts at a lower frequency (see Fig.~\ref{fig:ampli_freq}), which will compete with attenuation for the magnetic field present in the target tile.
In conclusion, based on Fig.\,\ref{fig:pairs} which summarize the results of hundreds of regions,
we did not find any overall amplitude variation for any mode between a flaring and a non-flaring region larger than $\sim$10\,\%.
However, if there is a smaller increase, we are not able to detect with our analysis.

As first proposed by \citet{Wolff_1972_1}, and more recently by \citet{Kumar17}, only some flares would increase the mode amplitude, and not necessarily the ones with largest peak X-ray flux. We look for this in our data. In the Weak and Strong Flares set, there are a dozen or more flaring regions with a relative amplitude difference from its paired region,
($\Delta\!A\!/\!A_{_Q}$)$_{\text{WF or SF}}$ - ($\Delta\!A\!/\!A_{_Q}$)$_{\text{WO}}$,
larger than 0.2 for at least 60\,\% of modes, but there are a similar number of flaring regions that have a difference $\la$ -0.2. 
This variation in the differences
is expected for a normal distribution with a standard deviation equal to 0.1.
Thus, we did not convincingly find an influence of any flare in our data set on the mode amplitude.
Given the diversity of magnetic configurations in and around the flaring or its comparison (non-flaring) region, the MAI gives a useful but incomplete characterisation.

\section{Conclusions}\label{sec:conclusions}

We analysed several factors that affect the observed amplitude of high-degree solar $p$-modes,
namely, line-of-sight, strong magnetic fields and gaps in the observations.
If not taken into account, the apparent changes in the observed mode amplitude could be misinterpreted as an effect of a flare.
Care is required not only in analysing the flaring region but also in the comparison one
that is used as the quiescent state to infer the change in the mode amplitude due to the flare in an ever-changing Sun.

We analysed the most intense flare observed in the space age, the so-called, Halloween flare,
applying ring-diagram analysis to MDI data.
Contrary to the results obtained by \citet{maurya2009}, we did not find any amplification
(larger than $\sim$\,5\,\%) in the inferred mode amplitude during the flare, whereas
\citet{maurya2009} reported an increase of more than 100\,\% in the mode amplitude during the flare.
This difference could be in part because they did not correct for
the line-of-sight effect as the observed region rotates in relation to the observer.
We found an increase of as much as 60\,\% in the mode amplitude during the flare, when not taking this into account.
On the other hand, we found that the dispersion in the fitted mode amplitudes increases as the duty cycle decreases, especially when it is lower than 80\,\%.

Next, we adopted a statistical approach and analysed all flares contained in the HEK database during more than four years at the high activity phase of Cycle 24 (2012\,--\,2016). The observed mode amplitudes were obtained by the HMI ring-diagram pipeline for the five-degree tiles. We divided the flares into two groups: flares with energy between 10\,--\,60 erg\,cm$^{-2}$~s$^{-1}$ and larger than 1200 \,cm$^{-2}$~s$^{-1}$.
Although we observed that the Strong Flares are associated with HMI tiles whose MAI has a median of 120\,G (twice the median for tiles with Weak Flares), 10\,\% of regions with Strong Flares have an MAI less than 30\,G (and 10\,G for Weak Flares).
Since the flares are a product of magnetic reconnection, they most often happen on, or in the vicinity of, strong magnetic fields.
In fact, we observed that the HMI tiles with small MAI associated with Strong Flares have in their vicinity (less than ~8$\degr$ away) regions with MAI $\sim$12 times larger than its MAI. 
The mode amplitude in a magnetically quiet tile is affected by a nearby active region by as much as 20\,\% in its observed amplitude, as shown by \cite{rabellosoares2016,rabellosoares2018}.
We see the effect of nearby active regions in the results of our analysis. 
This is another effect that must be taken into account when looking for an effect of a flare in the mode amplitudes. 
The influence of nearby active regions complicated matters further since the increase in mode amplitude, known as the acoustic halo, occurs at frequencies smaller than the acoustic cutoff frequency, at $\nu$ > 4.2 mHz \citep{rabellosoares2016}.

Taking all these effects into account, we found that the mode amplitudes do not vary on average more than 10\,\% of their amplitude between flaring and non-flaring regions. Thus, we conclude that if there were a change in the solar acoustic mode amplitudes due to the large energy released during a flare, the change would be small. We did not find any individual flaring region that has an increase in its mode amplitude outside the uncertainties in our analysis (i.e., 2.5 sigma = 25\,\% variation).

\section*{Acknowledgements}

We thank J. Leibacher for a detailed reading of the original manuscript and several comments that have greatly improved it.
This work was carried out with the support of
Fundação de Amparo à Pesquisa de Minas Gerais (FAPEMIG).
V.G.T. would like to acknowledge that: 
"This study was financed in part by the Coordenação de
Aperfeiçoamento de Pessoal de Nível Superior - Brasil (CAPES) - Finance Code 001".

The automated detection routines on the HEK database: \textit{SolarSoft}, \textit{SWPC}, and \textit{Feature Finding Team (FFT)} are provided by the Harvard-Smithsonian Center for Astrophysics (CfA/SAO), Space Weather Prediction Center (SWPC/NOAA), and Lockheed Martin Solar and Astrophysics Laboratory (LMSAL), respectively.
%
This research has made use of SunPy, an open-source and free community-developed solar data analysis package written in Python (HEK module).
MDI data is provided by the SOHO/MDI consortium. SOHO is a project of international cooperation between ESA and NASA. HMI data is provided by NASA/SDO and HMI science team. This work utilizes data obtained by the Global Oscillation Network Group (GONG) program, managed by the National Solar Observatory, which is operated by AURA, Inc. under a cooperative agreement with the National Science Foundation. The data were acquired by instruments operated by the Big Bear Solar Observatory, High Altitude Observatory, Learmonth Solar Observatory, Udaipur Solar Observatory, Instituto de Astrofíısica de Canarias, and Cerro Tololo Interamerican Observatory.

\section*{Data Availability}

The data underlying this article are available from the Joint Science Operations Center (http://jsoc.stanford.edu) and from the GONG Data Archive (https://gong.nso.edu).



\bibliographystyle{mnras}
\bibliography{rabellosoares_R3} 


\bsp	
\label{lastpage}
\end{document}